\documentclass[11pt]{article}

\usepackage{fullpage}
\usepackage[margin = 2.5cm]{geometry}

\usepackage{authblk}

\usepackage{amsmath, amssymb, amsthm,bbm}
\usepackage{algorithmic}
\usepackage[ruled]{algorithm}
\usepackage{graphicx}
\usepackage{color}
\usepackage[dvipsnames]{xcolor}

\usepackage[hyperindex,breaklinks]{hyperref}

\usepackage{nicefrac}

\newtheorem{theorem}{Theorem}[section]
\newtheorem{claim}[theorem]{Claim}
\newtheorem{corollary}[theorem]{Corollary}
\newtheorem{lemma}[theorem]{Lemma}
\newtheorem{definition}[theorem]{Definition}

\newcommand{\pr}{\mathbb{P}}
\newcommand{\E}{\mathbf{E}}

\newcommand{\opt}{\operatorname{OPT}}

\newcommand{\bbR}{\mathbb{R}}

\newcommand{\calC}{\mathcal{C}}

\newcommand{\calP}{\mathcal{P}}
\newcommand{\calQ}{\mathcal{Q}}

\newcommand{\calS}{\mathcal{S}}

\DeclareMathOperator*{\argmin}{arg\,min}
\DeclareMathOperator*{\argmax}{arg\,max}
\newcommand\floor[1]{\lfloor#1\rfloor}
\newcommand{\one}{\mathbbm{1}}
\newcommand\norm[1]{\Vert#1\Vert}

\title{Approximation Algorithm for Norm Multiway Cut\thanks{This work is supported by supported by NSF Awards CCF-1955351, CCF-1934931,  EECS-2216970, NSF CCF-1955173, CCF-1934843, and ECCS-2216899.}} 

\author[1]{Charlie Carlson} 
\author[2]{Jafar Jafarov} 
\author[3]{Konstantin Makarychev}
\author[2]{\authorcr Yury Makarychev}
\author[3]{Liren Shan}

\affil[1]{University of Colorado Boulder}
\affil[2]{Toyota Technological Institute at Chicago}
\affil[3]{Northwestern University}
\date{}

\begin{document}

\maketitle

\begin{abstract}

We consider variants of the classic Multiway Cut problem. Multiway Cut asks to partition a graph $G$ into $k$ parts so as to separate $k$ given terminals. 
Recently, Chandrasekaran and Wang (ESA 2021) introduced $\ell_p$-norm Multiway, a generalization of the problem, in which the goal is to minimize the $\ell_p$ norm of the edge boundaries of $k$ parts. We provide an $O(\log^{\nicefrac{1}{2}} n\log^{\nicefrac{1}{2}+1/p} k)$ approximation algorithm for this problem, improving upon the approximation guarantee of $O(\log^{\nicefrac{3}{2}} n \log^{\nicefrac{1}{2}} k)$  due to Chandrasekaran and Wang.

We also introduce and study Norm Multiway Cut, a further generalization of Multiway Cut. We assume that we are given access to an oracle, which answers certain queries about the norm. We present an $O(\log^{\nicefrac{1}{2}} n \log^{\nicefrac{7}{2}} k)$ approximation algorithm with a weaker oracle and an $O(\log^{\nicefrac{1}{2}} n \log^{\nicefrac{5}{2}} k)$ approximation algorithm with a stronger oracle. Additionally, we show that without any oracle access, there is no $n^{1/4-\varepsilon}$ approximation algorithm for every $\varepsilon > 0$ assuming the Hypergraph Dense-vs-Random Conjecture.
\end{abstract}

\newpage

\section{Introduction}
In this paper, we consider a variant of the classic combinatorial optimization problem, Minimum Multiway Cut.
Given an undirected graph $G=(V,E)$ with edge weights $w:E \rightarrow\mathbb{R}_{\geq 0}$ and $k$ terminals $t_1,\dots, t_k\in V$, the Minimum Multiway Cut problem asks to partition  graph $G$ into $k$ parts $P_1,\dots, P_k$ so that  $P_i$ contains terminal $t_i$. The Multiway Cut objective is to minimize the number or total weight of cut edges. For $k=2$, the problem is equivalent to the minimum $st$-Cut problem. Dahlhaus, Johnson,  Papadimitriou, Seymour, and Yannakakis 
proved that it is NP-complete and APX-hard for every $k > 2$~\cite{DJPSY94}. They also gave a simple combinatorial $(2-2/k)$-approximation algorithm. Later Călinescu, Karloff, and  Rabani~\cite{CKR98} showed how to obtain a \nicefrac{3}{2} approximation using linear programming. This result was improved in a series of papers
by Karger, Klein, Stein, Thorup, and  Young~\cite{KKSTY99}, Buchbinder, Naor and Schwartz~\cite{BNS13}, and Sharma and Vondrák~\cite{SV14} (see also~\cite{BSW17}). The currently best known approximation factor is $1.2965$~\cite{SV14}. The  best known LP integrality gap and Unique Games Conjecture hardness is $1.20016$ due to B{\'e}rczi,  Chandrasekaran, Kir{\'a}ly, and Madan~\cite{BCKM20}
(see also~\cite{AMM17,FK20,MNRS08}).

In 2004, Svitkina and Tardos~\cite{ST04} introduced the Min-Max Multiway Cut problem. In this problem, as before, we need to partition graph $G$ into $k$ parts $P_1,\dots, P_k$ so that each $P_i$ contains one terminal $t_i$. However, the objective function is different: Min-Max Multiway Cut asks to minimize the maximum of edge boundaries of sets $P_i$ i.e., $\text{minimize } \max_i \delta(P_i)$, where $\delta(P_i)$ is the total weight of edges crossing the cut $(P_i, V\setminus P_i)$. Svitkina and Tardos~\cite{ST04}  gave an $O(\log^3 n)$ approximation algorithm for the problem. Later, Bansal, Feige, Krauthgamer, Makarychev, Nagarajan, Naor, and Schwartz~\cite{BFK+11} provided an $O(\sqrt{\log n \log k})$-approximation algorithm.
Also, Ahmadi, Khuller, and Saha~\cite{AKS19} studied a related Min-Max Multicut problem.

Recently, Chandrasekaran and Wang~\cite{CW21} proposed a common generalization of the Min Multiway Cut and Min-Max Multiway Cut problems, which they called Minimum $\ell_p$-norm Multiway Cut. This problem asks to minimize the $\ell_p$ norm of the edge boundaries of parts $P_1,\dots, P_k$. In other words, the objective is to
\[
\textbf{minimize:\   }\Big(\sum_{i=1}^k \delta(P_i)^p\Big)^{1/p}.
\]
Note that this problem is equivalent to Min Multiway Cut when $p=1$ and to Min-Max Multiway Cut when $p=\infty$. Chandrasekaran and Wang~\cite{CW21} gave an $O(\log^{\nicefrac{3}{2}} n \log^{\nicefrac{1}{2}} k)$ approximation for the problem. Further, they proved that the problem is NP-hard for every $p \geq 1$ and $k \geq 4$. Moreover, it does not admit an $O(k^{1-1/p - \varepsilon})$-approximation for every $\varepsilon > 0$ assuming the Small Set-Expansion Conjecture; a natural convex program for the problem has the intgerality gap of $\Omega(k^{1-1/p})$.

In this paper, we provide an improved $O(\log^{\nicefrac{1}{2}} n \log^{\nicefrac{1}{2} + 1/p} k)$
approximation algorithm. We note that for $p=\infty$, our approximation guarantee matches the approximation of the algorithm due to Bansal et al.~\cite{BFK+11}.%
\footnote{Our algorithm is stated only for the case where $p$ is finite. However, we can solve an instance with $p=\infty$ by running the algorithm with $p = \log k$. Since $\|\cdot\|_{\log k}$ is within a constant factor of $\|\cdot\|_\infty$ for vectors in ${\mathbb R}^k$, this approach yields an $O(\sqrt{\log n} \log^{\nicefrac{1}{2} + \nicefrac{1}{\log k}}k)= O(\sqrt{\log n \log k})$-approximation.} For $p=2$, our approximation guarantee is $O(\log^{\nicefrac{1}{2}} n \log k)$, which is $\Theta(\log n/\sqrt{\log k})$ times better than the approximation guarantee of the algorithm due to  Chandrasekaran and Wang~\cite{CW21}.
We also consider variants of Multiway Cut with norms other than the $\ell_p$ norm.


\subsection{Our Results}

We now formally state our results. First, we present an approximation algorithm for the $\ell_p$-norm Multiway Cut problem. We show that our algorithm achieves an $O(\log^{\nicefrac{1}{2}} n \log^{\nicefrac{1}{2}+\nicefrac{1}{p}} k)$ approximation 
for every $p > 1$.

\begin{theorem}\label{thm:lp}
    There exists a polynomial-time randomized algorithm that given a graph with $n$ vertices, $k$ terminals, and $p>1$, finds an $O(\log^{\nicefrac{1}{2}} n \log^{\nicefrac{1}{2}+\nicefrac{1}{p}} k)$ approximation for $\ell_p$-norm Multiway Cut with high probability.
\end{theorem}

Further, we provide approximation algorithms for Norm Multiway Cut with an arbitrary monotonic norm,
a further generalization of $\ell_p$-norm Multiway Cut.
The monotonic norm is defined as follows.
\begin{definition}
\label{def:norm-monotonic}
A norm $\norm{\cdot}$ on $\bbR^d$ is monotonic if for any $x,y \in \bbR^d$ with $|x_i| \leq |y_i|$ for all $i \in [d]$, it holds $\norm{x} \leq \norm{y}$.
\end{definition}
We consider two oracles to the monotonic norm used in the Norm Multiway Cut: (1) minimization oracle; (2) ordering oracle. For a set $A \subseteq [d]$, let $ \one_A \in \{0,1\}^d$ denote the indicator vector of $A$, i.e., the $i$-th coordinate $(\one_A)_i=1$ if $i\in A$; otherwise, $(\one_A)_i=0$. 

\begin{definition}
\label{def:oracle-min}
Given a monotonic norm $\norm{\cdot}$ on $\bbR^d$, for any $i \in [d]$,  the minimization oracle efficiently finds a set $A_i \subseteq [d]$ that minimizes the norm of indicator vectors among all subsets with size $i$, i.e. 
\[A_i = \argmin_{A \subseteq [n], |A| = i} \norm{\one_{A}}.\]
\end{definition}

\begin{definition}
\label{def:oracle-order}
Given a monotonic norm $\norm{\cdot}$ on $\bbR^d$, for any vector $x \in \bbR^d$, the ordering oracle efficiently finds an ordering of the vector $x$ that minimizes the norm, i.e.
\[\pi_x = \argmin_{\pi \in S_d} \norm{x^\pi},\]
where $x^\pi$ denotes the ordering of $x$ regarding the permutation $\pi$.
\end{definition}

Assuming that they are given access to either a ``minimization oracle'' or a stronger ``ordering oracle'', our algorithms give $O(\log^{\nicefrac{1}{2}} n \log^{\nicefrac{7}{2}} k)$ and $O(\log^{\nicefrac{1}{2}} n \log^{\nicefrac{5}{2}} k)$ approximation, respectively.
We remark that the oracles only answer queries about the norm and, in particular, there is an ordering oracle for the $\ell_p$-norm, weighted $\ell_p$-norm,  and many other natural norms.
Thus, our result implies an $O(\log^{\nicefrac{1}{2}} n \log^{\nicefrac{5}{2}} k)$ approximation for \textit{weighted} $\ell_p$-norm Multiway Cut. 
We prove the following theorems in Section~\ref{sec:Norm Multiway Cut}.

\begin{theorem}\label{thm:minimization-oracle}
    There exists a polynomial-time  algorithm that for every monotonic norm with a minimization oracle, given a graph with $n$ vertices and $k$ terminals, finds an $O(\log^{\nicefrac{1}{2}} n \log^{\nicefrac{7}{2}} k)$ approximation for the Norm Multiway Cut with high probability.
\end{theorem}

\begin{theorem}\label{thm:ordering-oracle}
    There exists a polynomial-time algorithm that for every monotonic norm with an ordering oracle, given a graph with $n$ vertices and $k$ terminals, finds an $O(\log^{\nicefrac{1}{2}} n \log^{\nicefrac{5}{2}} k)$ approximation for the Norm Multiway Cut with high probability.
\end{theorem}

Finally, we show that the problem becomes very hard if we are not given access to a norm minimization oracle. The proof is given in the Section~\ref{sec:hardness}.

\begin{theorem}
Consider the Norm Multiway Cut problem with a monotonic norm. Assume that the norm is given by a formula (in particular, we can easily compute the value of the norm; however, we are not given a minimization oracle for it).
Then, assuming the Hypergraph Dense-vs-Random Conjecture,  there is no polynomial-time algorithm for Norm Multiway Cut with approximation factor $\alpha(n) \leq n^{1/4-\varepsilon}$ for every $\varepsilon >0$.
\end{theorem}

\subsection{Proof Overview}

We first describe our algorithm for the $\ell_p$-norm Multiway Cut. Our algorithm consists of three procedures: (1) covering procedure, (2) uncrossing procedure, and (3) aggregation procedure. 

In the covering procedure, we generate a collection of subsets of the vertex set, $\calS = \{S_1,S_2,\cdots,S_m\}$. We generate these sets iteratively by using a bi-criteria approximation algorithm for Unbalanced Terminal Cut by Bansal et al.~\cite{BFK+11} and a multiplicative weight update method. See Section~\ref{sec:covering} and Algorithm~\ref{alg:covering} for details.  Each set in $\calS$ contains at most one terminal. These sets are not disjoint. While, these sets cover the entire graph, which means the union of all sets in $\calS$ contains all vertices. The number of sets in $\calS$ is at most $O(k\log n)$. We show that the $\ell_p$ norm and $\ell_1$ norm of the edge boundaries of sets in $\calS$ is at most $O(\log^{1/p} n \cdot\alpha) \opt$ and  $O(\log n \cdot k^{1-1/p} \cdot\alpha) \opt$ respectively, where $\alpha = \sqrt{\log n\log k}$ and $\opt$ is the cost of the optimal solution. This covering procedure follows the approach by Bansal et al.~\cite{BFK+11} for Min-Max Multiway Cut. Chandrasekaran and Wang~\cite{CW21} also use a similar covering procedure for $\ell_p$-norm Multiway Cut. Their algorithm finds a cover $\calS$ that satisfies the above properties except for the $\ell_1$ norm bound. We get this $\ell_1$ norm bound on the edge boundaries of sets in $\calS$ by picking proper measure constraints for the Unbalanced Terminal Cut algorithm. This $\ell_1$ norm bound is important in the aggregation procedure to get an improved approximation.    

\enlargethispage*{0.075cm}

Note that the sets in $\calS$ are not disjoint. We use the uncrossing procedure to create a partition of the graph with at most $O(k\log k)$ sets. Our uncrossing procedure first sample $O(k\log k)$ sets from $\calS$ uniformly at random. Then, we run an iterative uncrossing process given by Bansal et al.~\cite{BFK+11} over sampled sets until all sets are disjoint and have small boundaries. We show that all sampled sets cover almost the entire graph. The set of uncovered vertices does not contain terminals and has a small boundary with high probability. Next, we use the aggregation procedure to merge these $O(k\log k)$ sets into a $k$ partition. We assign $k$ sets containing one terminal to $k$ parts. For other sets without terminals, we assign them to $k$ parts almost uniformly such that each part has almost the same $\ell_1$ norm over assigned sets. After the uncrossing procedure, the $\ell_p$ norm and $\ell_1$ norm of edge boundaries is at most $O(\log^{1/p} k \cdot\alpha) \opt$ and $O(k^{1-1/p} \cdot\alpha) \opt$ respectively. We upper bound the sets containing one terminal and the sets with the largest edge boundary in each part by the above $\ell_p$ norm bound. For the remaining sets, by the $\ell_1$ norm bound and the uniform assignment, we upper bound the $\ell_p$ norm for these sets by $O(\alpha)\opt$. Chandrasekaran and Wang~\cite{CW21} achieve an $O(\log n \cdot \alpha)$ where the $O(\log n)$ factor is due to their aggregation procedure. 
We use the $\ell_1$ norm bound in the covering procedure and a new aggregation procedure to reduce $O(\log n)$ extra factor to $O(\log^{1/p} n)$. We use the sampling in the uncrossing procedure to further reduce the extra factor from $O(\log^{1/p} n)$ to $O(\log^{1/p} k)$. 

We now describe our algorithm for Norm Multiway Cut. We use the same framework with covering, uncrossing, and aggregation procedures. While, unlike the $\ell_p$ norm, the general monotonic norm may not be permutation invariant. For each terminal, we first compute a minimum cut that separates this terminal from other terminals. Then, we can remove all terminals and assign the remaining vertices freely among $k$ parts.
We mainly use a bucketing idea to modify our algorithm. 
We partition $k$ coordinates into $\log_2 k$ buckets with exponentially increasing size $2^i$ such that the coordinates with large boundaries in the optimal solution are assigned to small buckets. 
Differing from the previous covering procedure, the new covering procedure uses the Unbalanced Terminal Cut algorithm with parameters related to each bucket. 
The cover $\calS$ contains $O(2^i\log n\log k)$ sets in each bucket $i$. The boundary of every set in each bucket is relatively small, at most $O(\alpha)$ times the boundary of the optimal part in this bucket. We then run the uncrossing and aggregation procedure to create a multiway cut. We still sample each set in $\calS$ with probability $\log_2 k/\log_2 n$. Thus, we have $O(2^i \log^2 k)$ sets in each bucket $i$ after the uncrossing procedure.
For bucket $0\leq i \leq \log_2 k$, we find a set of $2^i$ coordinates $I_i \in [k]$ that minimizes the norm of the indicator vector through the minimization oracle. We then assign $O(2^i \log^2 k)$ sets in each bucket to coordinates in $I_i$ such that each coordinate has $O(\log^2 k)$ sets in bucket $i$. Thus, we achieve an $O(\log^2 k \cdot \alpha)$ approximation for each bucket. Since these sets of coordinates $I_i$ may overlap, we lose an additional $O(\log k)$ factor for $\log_2 k$ buckets. Suppose we have a stronger oracle that finds the best ordering for any given vector that minimizes the norm. Then, we provide an assignment for each bucket to avoid the large overlapping among buckets. Therefore, we avoid the extra $O(\log k)$ factor loss due to the overlapping.  

\section{\texorpdfstring{$\ell_p$}{lp}-norm Multiway Cut}
In this section, we present our algorithm for $\ell_p$-norm Multiway Cut. We prove the following theorem. Our algorithm consists of three parts: covering procedure, uncrossing procedure, and aggregation procedure. We describe and analyze the covering procedure in Section~\ref{sec:covering}, the uncrossing and aggregation procedures in Section~\ref{sec:aggregation}.

\newtheorem*{thm:lp}{Theorem \ref{thm:lp}}
\begin{thm:lp}
    There exists a polynomial-time randomized algorithm that given a graph with $n$ vertices and $k$ terminals, and $p>1$, finds an $O(\log^{\nicefrac{1}{2}} n \log^{\nicefrac{1}{2}+\nicefrac{1}{p}} k)$ approximation for the $\ell_p$-norm Multiway Cut with high probability.
\end{thm:lp}

\subsection{Covering Procedure}\label{sec:covering}
We first present and analyze a covering procedure in our algorithm.  The covering procedure takes a undirected graph $G=(V,E)$ with edge weights $w:E \rightarrow\mathbb{R}_{\geq 0}$ and $k$ terminals $T = \{t_1,\dots, t_k\} \subset V$ as input and outputs a collection of sets $\calS = \{S_1,\dots, S_m\}$ where $S_i \subset V$ for all $i$. All sets $S_i \in \calS$ covers the entire graph, $\bigcup_{i=1}^m S_i = V$. Each set $S_i\in \calS$ contains at most one terminal. 
For each subset $S \subseteq V$, we use $\partial(S) = E(S,V\setminus S)$ to denote all edges crossing the cut $(S,V\setminus S)$. We use $\delta(S) = \sum_{e \in \partial(S)} w(e)$ to denote the edge boundary of set $S$, which is the total weight of all edges crossing $(S,V\setminus S)$.
We prove the following upper bounds on the $\ell_1$-norm and $\ell_p$-norm of the edge boundaries of these sets in $\calS$, which is crucial for our approximation guarantee. 

\begin{lemma}\label{lem:set_cover}
Given a graph $G=(V,E)$ with $n$ vertices and  $k$ terminals $T \subset V$, the covering procedure shown in Algorithm~\ref{alg:covering} returns $m = O(k\log n)$ sets $\calS = \{S_1,\dots, S_m\}$ that satisfies
\begin{enumerate}
    \item $|S_i \cap T| \leq 1$ for all $i \in [m]$,
    \item $\bigcup\limits_{i=1}^m S_i = V$,
    \item $\sum\limits_{t=1}^{m}\delta(S_t)^p\leq \log n\cdot O(\alpha^p) \cdot \opt^p $,
    \item $\sum\limits_{t=1}^{m}\delta(S_t)\leq \log n\cdot O(\alpha)\cdot k^{1-\nicefrac{1}{p}} \cdot \opt$,
\end{enumerate}
where $\alpha = \sqrt{\log n \log k}$ and $\opt$ is the objective value of the optimal $\ell_p$-norm Multiway Cut.
\end{lemma}

Our algorithm relies on an intermediate problem, Unbalanced Terminal Cut that we introduce now.

\begin{definition}[Unbalanced Terminal Cut]
The input to this problem is a tuple $\langle G, w, \mu, \rho, T\rangle$, where $G=(V,E)$ is a graph with edge weights $w:E\rightarrow\mathbb{R}_{\geq 0}$, a measure $\mu:V\rightarrow\mathbb{R}_{\geq 0}$, a parameter $\rho\in (0,1]$, and a set of terminals $T$. The goal is to find $S\subseteq V$ of minimum cost $\delta(S)$ satisfying:
\begin{enumerate}
    \item $\vert S\cap T \vert \leq 1$,
    \item $\mu(S)\geq \rho\cdot \mu(V)$.
\end{enumerate}
\end{definition}

Bansal, Feige, Krauthgamer, Makarychev, Nagarajan, Naor, and Schwartz \cite{BFK+11} gave a bi-criteria approximation algorithm for Unbalanced Terminal Cut that we state in the following theorem.
\begin{theorem}\label{thm:covering:UTC}
There exists a polynomial-time algorithm that given an instance $\langle G, w, \mu, \rho, T\rangle$ of Unbalanced Terminal Cut, finds a set $S\subseteq V$ satisfying $\vert S\cap T \vert\leq 1$, $\mu(S)\geq \Omega(\rho)\cdot \mu(V)$, and $\delta(S)\leq \alpha\cdot \opt_{\langle G, w, \mu, \rho, T\rangle}$ where $\alpha = O\left(\sqrt{\log n \log\nicefrac{1}{\rho}}\right)$.
\end{theorem}

Our covering procedure relies on the multiplicative weights update method and is inspired by the algorithm in~\cite{BFK+11}. It initializes the measure of each vertex to one. At each iteration $t$, the algorithm guesses the measure $\mu_t(P_i^*)$ of a particular set $P_i^*$ in an optimal solution and computes $S_t$ of measure $\mu_t(S_t)\approx \mu_t(P_i^*)$ using the Unbalanced Terminal Cut algorithm in Theorem~\ref{thm:covering:UTC}. The existence of such a $P_i^*$ is shown in Lemma~\ref{lmm:covering:opt_exists}. Once $S_t$ is computed, the algorithm decreases the measure of the vertices covered by $S_t$ by a factor of $2$. The algorithm terminates when the total measure of vertices is less than $\nicefrac{1}{n}$.

We guess $\mu_t(P_i^*)$ as follows: For any set $S\subseteq V$, its measure $\mu_t(S)$ lies in the range $[\mu_t(u),n\cdot\mu_t(u)]$, where $u=\arg\max_{v\in S}\mu_t(v)$ is the heaviest vertex in $S$. Thus $\mu_t(P_i^*)$ can be well approximated by the set $A = \left\{ 2^i\cdot \mu_t(v): v\in V,i=0,\dots,\lfloor \log_2 n \rfloor \right\}$ of size $O(n\log n)$. For each candidate $a\in A$ we compute a set $S(a)$ using the Unbalanced Terminal Cut algorithm with a parameter $a$ and choose $S_t=\arg\min_{a\in A}\delta(S(a))$ with the smallest cost. We give a pseudo-code for this algorithm in Algorithm~\ref{alg:covering}. We remark that one can think of this algorithm as of multiplicative weight update algorithm for solving a covering LP with constraints from Lemma~\ref{lem:set_cover}.

\begin{algorithm}[t]
   \caption{Covering Procedure}
   \label{alg:covering}
\begin{algorithmic}
   \STATE Set $t=1$, and $\mu_1(v)=1$ for all $v\in V$. Let $\calS = \varnothing$. \\
   \WHILE{$\sum\limits_{v\in V}\mu_t(v)\geq \frac{1}{n}$}
   \STATE Let $P_i^*$ be a set as stated in Lemma~\ref{lmm:covering:opt_exists}.
   \STATE Guess $\mu_t(P^*_i)$.
   \STATE Let $S_t\subseteq V$ be the solution for Unbalanced Terminal Cut instance $\langle G,w,\mu_t,\max\{\frac{1}{2k},\frac{\mu_t(P^*_i)}{\mu_t(V)}\},T \rangle$.
   \STATE Let $\calS = \calS \cup \{S_t\}$.
   \FOR{$v\in V$}
\STATE Set $\mu_{t+1}(v) = 
     \begin{cases}
       \nicefrac{\mu_t(v)}{2}, &\quad\text{if } v\in S_t,\\
       \mu_t(v), &\quad\text{if } v \not\in S_t.
     \end{cases}$
   \ENDFOR
   \STATE Set $t = t+1$
   \ENDWHILE
   \RETURN $\calS$.
\end{algorithmic}
\end{algorithm}

We then analyze this covering algorithm in Algorithm~\ref{alg:covering}. Let $\calS=\{S_1,\cdots, S_m\}$ denote the collection of $m$ sets output by Algorithm~\ref{alg:covering}. By Theorem~\ref{thm:covering:UTC}, every set $S_i$ contains at most one terminal.  First, for any fixed vertex $v\in V$, we give a lower bound on the number of sets containing $v$.

\begin{claim}\label{norm:claim:lower_bound_on_frequency}
For a vertex $v\in V$, let $N_v=\vert\{S_t\vert\;v\in S_t\}\vert$ denote the number of sets containing $v$. Then $N_v\geq \Omega(\log n)$.
\end{claim}
\begin{proof}
Recall that initially $\mu_1(v) = 1$ and after iteration $m$ its measure becomes $\mu_{m+1}(v)=\left(\frac{1}{2}\right)^{N_v}$. Due to the stopping condition of our algorithm we have $\mu_{m+1}(V)<\frac{1}{n}$. Thus, $\frac{1}{2^{N_v}} < \frac{1}{n}$ and the claim follows.
\end{proof}

Next, we bound the number of sets in $\calS$.
In the following claim we give an upper bound on the total normalized measure of the sets produced by our algorithm.
\begin{claim}\label{clm:covering:total_measure_bound}
$\sum\limits_{t=1}^{m}\frac{\mu_t(S_t)}{\mu_t(V)}\leq 4\ln n+ 1$.
\end{claim}
\begin{proof}
Observe that the total measure at iteration $t$ can be described as follows:
\begin{align*}
    \mu_t(V) = \mu_{t-1}(V)-\frac{\mu_{t-1}(S_{t-1})}{2} = \mu_{t-1}(V)\cdot\left(1-\frac{\mu_{t-1}(S_{t-1})}{2\mu_{t-1}(V)}\right).
\end{align*}
Since $\mu_{m}(V)\geq \frac{1}{n}$, we have
\begin{align*}
\frac{1}{n}\leq \mu_{m}(V)  = \mu_1(V)\cdot \prod\limits_{t=1}^{m-1}\left(1-\frac{\mu_{t}(S_{t})}{2\mu_{t}(V)}\right)\leq n \cdot \prod\limits_{t=1}^{m-1}e^{-\frac{\mu_{t}(S_{t})}{2\mu_{t}(V)}} = n\cdot e^{-\frac{1}{2}\cdot\sum\limits_{t=1}^{m-1}\frac{\mu_{t}(S_{t})}{\mu_{t}(V)}},
\end{align*}
which implies
$  \sum\limits_{t=1}^{m-1}\frac{\mu_{t}(S_{t})}{\mu_{t}(V)} \leq 4\ln n 
$.
Since $\frac{\mu_m(S_m)}{\mu_m(V)}\leq 1$, we get the desired result.
\end{proof}

We obtain an upper-bound on the number of sets in $\calS$ which immediately follows from Claim~\ref{clm:covering:total_measure_bound} and the fact that $\mu_t(S_t)\geq \Omega ({1}/{k})\mu_t(V)$ for all $t$.

\begin{corollary}\label{col:number of cover sets}
The cover $\calS$ returned by Algorithm~\ref{alg:covering} contains $m=  O(k\log n)$ sets.
\end{corollary}

We prove the existence of a set in an optimal solution with large measure and small cut value.
\begin{lemma}\label{lmm:covering:opt_exists}
Let $\calP^{*}=\left(P_1^*,\dots,P_k^*\right)$ be an optimal solution to an $\ell_p$-norm Multiway Cut instance and let $\opt$ denote the $\ell_p$-norm of $\calP^*$. For any measure $\mu:V\rightarrow \mathbb{R}_{\geq 0}$ on vertices such that $\mu(V) \not= 0$, there exists an $i\in [k]$ such that the following three conditions hold:
\begin{enumerate}
    \item $\delta(P_i^*)^p\leq 5\cdot \opt^p \cdot \frac{\mu(P_i^*)}{\mu(V)}$
    \item $\delta(P_i^*)\leq 5k^{1-\nicefrac{1}{p}}\cdot \opt \cdot \frac{\mu(P_i^*)}{\mu(V)} $
    \item $\mu(P^*_i)\geq \frac{\mu(V)}{2k}$
\end{enumerate}
\end{lemma}

\begin{proof}
    Let 
    \begin{equation*}
    J = \{j \in [k]:  \delta(P_j^*)^p\leq 5\cdot \opt^p \cdot \nicefrac{\mu(P_j^*)}{\mu(V)}, \delta(P_j^*)\leq 5k^{1-\nicefrac{1}{p}}\cdot \opt \cdot \nicefrac{\mu(P_j^*)}{\mu(V)} \}
    \end{equation*}
    be the indices of sets in $\calP^*$ that satisfies conditions 1 and 2 in Lemma~\ref{lmm:covering:opt_exists}. It is sufficient to show that $\sum_{j \in [k] \setminus J} \mu(P_j^*) < \mu(V)/2$. If $\sum_{j \in [k] \setminus J} \mu(P_j^*) < \mu(V)/2$, then there exists a $j \in J$ such that $\mu(P_j^*) \geq \nicefrac{\mu(V)}{2k}$, which implies this set $P_j^*$ satisfies all three conditions.

    We now show that $\sum_{j \in [k] \setminus J} \mu(P_j^*) < \mu(V)/2$. Let $\bar J_1 = \{j \in [k]\setminus J: \delta(P_j^*)^p > 5\cdot \opt^p \cdot \nicefrac{\mu(P_j^*)}{\mu(V)}\}$ be the indices of sets $P_j^*$ that does not satisfy condition 1. 
    Let $\bar J_2 = \{j \in [k]\setminus J: \delta(P_j^*) > 5k^{1-\nicefrac{1}{p}}\cdot \opt \cdot \nicefrac{\mu(P_j^*)}{\mu(V)}\}$ be the indices of sets $P_j^*$ that does not satisfy condition 2. Note that $[k]\setminus J = \bar J_1 \cup \bar J_2$. Then, we have
    \begin{align*}
    \sum_{j \in [k]\setminus J} \mu(P_j^*) &\leq \sum_{j \in \bar J_1} \mu(P_j^*) + \sum_{j \in \bar J_2} \mu(P_j^*) \\
    &\leq \sum_{j \in \bar J_1} \mu(V) \cdot\frac{\delta(P_j^*)^p}{5\opt^p} + \sum_{j \in \bar J_2} \mu(V)\cdot \frac{\delta(P_j^*)}{5k^{1-\nicefrac{1}{p}}\opt}\\
    &\leq \mu(V)\cdot \sum_{j \in [k]} \left( \frac{\delta(P_j^*)^p}{5\opt^p} + \frac{\delta(P_j^*)}{5k^{1-\nicefrac{1}{p}}\opt} \right).
    \end{align*}
    Since $\calP^*$ is a partition with an optimal cost, we have
    \begin{align*}    \sum\limits_{i=1}^{k}\delta(P^*_i)^p = \opt^p.
    \end{align*}
    Similarly, we have
    \begin{align*}
    k^{-\nicefrac{1}{p}}\cdot \opt =\left( \sum\limits_{i=1}^{k}\frac{1}{k}\cdot\delta(P^*_i)^p\right)^{\nicefrac{1}{p}}\geq \frac{1}{k}\sum\limits_{i=1}^{k}\delta(P^*_i),
    \end{align*}
    where the inequality follows from Jensen's inequality. 
    Thus, we have
    \begin{equation*}
    \sum_{j \in [k]\setminus J} \mu(P_j^*) \leq \mu(V)\cdot \sum_{j \in [k]} \frac{\delta(P_j^*)^p}{5\opt^p} + \frac{\delta(P_j^*)}{5k^{1-\nicefrac{1}{p}}\opt} \leq \mu(V) \cdot \frac{2}{5} < \frac{\mu(V)}{2}.
    \end{equation*}
\end{proof}

We now prove the main lemma in this section. Specifically, we give two upper-bounds on the $\ell_1$-norm and $\ell_p$-norm of the cut values of the sets produced by the covering procedure in Algorithm~\ref{alg:covering}, respectively.

\begin{proof}[Proof of Lemma~\ref{lem:set_cover}]
We already show the number of sets in $\calS$ is at most $m = O(k\log n)$ in Corollary~\ref{col:number of cover sets}. By Theorem~\ref{thm:covering:UTC} and Claim~\ref{norm:claim:lower_bound_on_frequency}, we have every $S_i$ contains at most one terminal and all sets in $\calS$ covers the entire graph. Thus, it is sufficient to prove the two bounds on the $\ell_1$-norm and $\ell_p$-norm of the cut values of the sets in $\calS$ as shown in Conditions 3 and 4 in the lemma.

Due to Lemma~\ref{lmm:covering:opt_exists} at each iteration $t$, there exists a set $P^*_i $ in an optimal solution with a measure $\mu_t(P^*_i)\geq \frac{\mu_t(V)}{2k}$ such that
\[
    \delta(P^*_i)\leq 5\cdot \min\left\{\left(\frac{\mu_{t}(P^*_i)}{\mu_t(V)}\right)^{\nicefrac{1}{p}}, k^{1-\nicefrac{1}{p}}\cdot \frac{\mu_{t}(P^*_i)}{\mu_t(V)}\right\}\cdot \opt.
\]
Thus, at each iteration $t$ of Algorithm~\ref{alg:covering}, we have
\[
\delta(S_t)\leq O(\alpha)\cdot \min\left\{\left(\frac{\mu_{t}(P^*_i)}{\mu_t(V)}\right)^{\nicefrac{1}{p}}, k^{1-\nicefrac{1}{p}}\cdot \frac{\mu_{t}(P^*_i)}{\mu_t(V)}\right\}\cdot \opt.
\] 
Note that each set $S_t$ is computed by the Unbalanced Terminal Cut algorithm in Theorem~\ref{thm:covering:UTC}. Thus, we have $\mu_t(S_t)\geq \Omega(\mu_t(P^*_i))$.
Since $\mu_t(S_t)\geq \Omega(\mu_t(P^*_i))$ holds, we obtain (1) $\delta(S_t)^p\leq O(\alpha^p) \cdot \opt^p\cdot \frac{\mu_{t}(S_t)}{\mu_t(V)} $; and (2) $\delta(S_t)\leq O(\alpha)\cdot k^{1-\nicefrac{1}{p}} \cdot \opt \cdot \frac{\mu_{t}(S_t)}{\mu_t(V)}$. These provide
\begin{align*}
  \sum\limits_{t=1}^{m}\delta(S_t)^p &\leq O(\alpha^p) \cdot \opt^p\cdot \sum\limits_{t=1}^{m}\frac{\mu_{t}(S_t)}{\mu_t(V)}, \\ \sum\limits_{t=1}^{m}\delta(S_t) &\leq O(\alpha)\cdot k^{1-\nicefrac{1}{p}} \cdot \opt \cdot \sum\limits_{t=1}^{m}\frac{\mu_{t}(S_t)}{\mu_t(V)}.
\end{align*}
By Claim~\ref{clm:covering:total_measure_bound}, we have $\sum\limits_{t=1}^{m}\frac{\mu_{t}(S_t)}{\mu_t(V)} = O(\log n)$.
Then, we get the desired upper bounds on $\sum\limits_{t=1}^{m}\delta(S_t)^p$ and $\sum\limits_{t=1}^{m}\delta(S_t)$.
\end{proof}
\subsection{Uncrossing and Aggregation Procedures}\label{sec:aggregation}
In this section, we provide procedures that transform the cover of the graph $\calS$ produced by the covering procedure into a partition of the graph $\calP = \{P_1,P_2,\dots,P_k\}$. Each set $P_i$ in $\calP$ contains exactly one terminal in $T$. With a positive probability, this solution is an $O(\log^{\nicefrac{1}{2}} n \log^{\nicefrac{1}{2}+\nicefrac{1}{p}} k)$ approximation for the $\ell_p$-Norm Multiway Cut.

\begin{theorem}\label{thm:Lp-multiway-cut}
Given a graph $G=(V,E)$ and $k$ terminals $T\subset V$, there exists a polynomial-time algorithm that returns a partition of the graph $\calP = \{P_1,P_2.\dots, P_k\}$ such that with probability at least $3/4-1/k$
\begin{enumerate}
    \item $|P_i \cap T| = 1$ for all $i \in [k]$,
    \item $\left(\sum\limits_{i=1}^{k}\delta(P_i)^p\right)^{\nicefrac{1}{p}}\leq  O(\log^{\nicefrac{1}{2}} n \log^{\nicefrac{1}{2}+\nicefrac{1}{p}} k) \cdot \opt$.
\end{enumerate}
\end{theorem}

Note that the sets in the cover $\calS$ are not disjoint. We first use the uncrossing procedure to generate a $m'=O(k\log k)$ partition of the graph $\calP'= \{P'_1,P'_2,\dots,P'_{m'}\}$ from the cover $\calS$ produced by the covering procedure. We sample $O(k\log k)$ sets from $\calS$ uniformly at random. These sampled sets cover a large fraction of the graph. Then, we generate disjoint sets from these sampled sets by using the uncrossing step in \cite{BFK+11}. The uncrossing procedure is shown in Algorithm~\ref{alg:uncrossing}. We then merge these sets in $\calP'$ to get a $k$-partition $\calP$ using the aggregation procedure in Algorithm~\ref{alg:aggregating}.  

In the aggregation procedure, we assign all sets in $\calP'$ into $k$ parts to get a $k$-partition. Since $\calP'=\{P'_1,P'_2,\dots,P'_{m'}\}$ is a partition of the graph and each set $P'_i$ contains at most one terminal, there are exactly $k$ sets containing one terminal in $\calP'$. Suppose $P'_1,P'_2,\cdots,P'_k$ are these sets containing one terminal. We initially assign these sets $P'_1,P'_2,\cdots,P'_k$ to $k$ parts $P_1,P_2,\dots,P_k$. 
Let $\calQ = \calP'\setminus\{P_1,P_2,\cdots, P_k\}$ be the sets in $\calP'$ that does not contain any terminals.
We assign all sets in $\calQ$ into $k$ parts in a round-robin approach. 
We sort the sets in $\calQ$ by the cut values in descending order and denote it by $\calQ =\{Q_1,Q_2,\dots, Q_{m'-k}\}$. 
We then partition all sets in $\calQ$ into $k$ buckets $\calQ_1,\dots,\calQ_k$ as follows. Consider every $k$ consecutive sets $\{Q_{jk+1},Q_{jk+2},\cdots,Q_{(j+1)k}\}$ in $\calQ$ for $0\leq j \leq \lfloor \nicefrac{m'-k}{k} \rfloor$. If $jk+i > n$ for $j = \lfloor \nicefrac{m'-k}{k} \rfloor$ and some $i \in [k]$, then let $Q_{jk+i} = \varnothing$. For every $i \in [k]$, we assign the set $Q_{jk+i}$ to the bucket $\calQ_i$. Finally, we assign each bucket $\calQ_i$ to part $P_i$ and set $P_i = P_i \cup (\bigcup_{Q_j \in \calQ_i} Q_j)$.

\begin{algorithm}
    \caption{Uncrossing Procedure}
    \label{alg:uncrossing}
\begin{algorithmic}
    \STATE Sample $m''-1=12k\ln k$ sets $\calS' = (S'_1,S'_2,\dots,S'_{m''-1})$ from $\calS$ uniformly at random.
    \STATE Sort sets in $\calS'$ in a random order. \STATE Set $P'_i = S'_i \setminus \cup_{j<i} S'_j$ for all $i =1,2\dots,m''-1$.
    \WHILE{there exists a set $P'_i$ such that $\delta(P'_i) > 2\delta(S'_i)$}
    \STATE Set $P'_i = S'_i$ and for all $j\neq i$, $P'_j = P'_j \setminus S'_i$.
    \ENDWHILE
    \STATE Set the set $P'_{m''} = V\setminus \cup_{i=1}^{m''-1} P'_i$.
    \RETURN all non-empty sets $P'_i$.
\end{algorithmic}
\end{algorithm}
\begin{algorithm}
    \caption{Aggregation Procedure}
    \label{alg:aggregating}
\begin{algorithmic}
    \STATE Set $\calP = \{P'_i \in \calP' : P'_i \cap T \neq \varnothing\} = \{P_1,P_2,\dots,P_k\}$.
    \STATE Set $\calQ = \calP' \setminus \calP$.
    \STATE Sort the sets in $\calQ = \{Q_1,\dots,Q_{m'-k}\}$ by the cut value in descending order.
    \STATE Partition the sets in $\calQ$ into $k$ buckets $\calQ_1,\cdots,\calQ_k$, where $$\calQ_i = \{Q_j \in \calQ: (j-1)\mod k=i-1 \}.$$
    \STATE Set $P_i = (\bigcup_{Q_j \in \calQ_i} Q_j) \cup P_i$ for all $i=1,\dots,k$.
    \RETURN all sets $P_1,P_2,\dots,P_k$.
\end{algorithmic}
\end{algorithm}

We first prove the following lemma on the partition $\calP'$ returned by the uncrossing procedure.
\begin{lemma}\label{lem:uncrossing}
Let $\calS$ denote the collection of sets produced by the covering procedure for a graph $G=(V,E)$ and $k$ terminals $T$. Given $\calS$ as input, the uncrossing procedure as shown in Algorithm~\ref{alg:uncrossing} generates a $m' = O(k\log k)$ partition of the graph $\calP' =\{P'_1,P'_2,\dots,P'_{m'}\}$ such that with probability at least $3/4-1/k$
\begin{enumerate}
    \item $|P'_i \cap T| \leq 1$ for all $i\in [m']$,
    \item $\sum\limits_{i=1}^{m'}\delta(P'_i)^p \leq  O(\log k\cdot \alpha^p) \cdot \opt^p $,
    \item $\sum\limits_{i=1}^{m'}\delta(P'_i) \leq  O(k^{1-\nicefrac{1}{p}} \cdot \alpha)  \cdot \opt$,
\end{enumerate}
where $\alpha = \sqrt{\log n \log k}$.
\end{lemma}

\begin{proof}
We consider all sets $P'_1,P'_2,\dots,P'_{m''}$ generated in the uncrossing procedure (Algorithm~\ref{alg:uncrossing}), including those empty sets that are not returned. If the set $P'_i$ is empty, we take $\delta(P'_i) = 0$. It is easy to see that these sets $P'_1,P'_2,\dots,P'_{m''}$ are disjoint, and $\bigcup_{i=1}^{m''} P'_i = V$.

We first show that Algorithm~\ref{alg:uncrossing} terminates in polynomial time. In graph $G$, we assume that the ratio between the largest non-infinite edge weight $w_{max}$ and the smallest non-zero edge weight $w_{min}$ is at most $w_{max}/w_{min} \leq n^2/\varepsilon$ for a small constant $\varepsilon > 0$. If the graph does not satisfy this assumption, then we transform it into an instance satisfying this condition as follows. We guess the largest weight of the cut edge in the optimal solution, denoted by $W$. There are at most $O(n^2)$ different edge weights. Then, we construct a new graph $G'$ with the same vertex set $V$ and edge set $E$. For every edge $e\in E$, we assign its weight $w'(e)$ in $G'$ to be $w(e)$ if $\varepsilon W/n^2\leq w(e) \leq W$, $w'(e) = 0$ if $w(e) < \varepsilon W/n^2$, and $w'(e) = \infty$ if $w(e) \geq W$. Thus, the new graph $G'$ satisfies the assumption that $w_{max}/w_{min} \leq n^2/\varepsilon$. Let $\opt'$ be the optimal value of $\ell_p$ multiway cut on graph $G'$. We know that $\opt' \leq \opt$ since the optimal multiway cut on graph $G$ has a smaller value on graph $G'$. Suppose we find an $\alpha$-approximation for $\ell_p$ multiway cut on graph $G'$. Then, the same partition on the original graph $G$ has an objective value at most $\alpha \cdot \opt' + \varepsilon W \leq (\alpha+\varepsilon)\opt$. Hence, this $\alpha$-approximation solution on $G'$ provides an $(\alpha+\varepsilon)$-approximation on $G$.

Consider any iteration of Algorithm~\ref{alg:uncrossing}. Let $P'_i$ be the partition of $V$ before the current uncrossing iteration. Suppose we pick a set $P'_i$ such that $\delta(P'_i) > 2\delta(S'_i)$. For any two subsets $A,B \subseteq V$, we use $\delta(A,B)$ to denote the total weight of edges crossing $A$ and $B$. Then, we have the $\ell_1$-norm of the cut values after this iteration is 
\begin{align*}
    \delta(S'_i) + \sum_{j\neq i} \delta(P'_j\setminus S'_i)  &\leq \delta(S'_i) + \sum_{j\neq i} \delta(P'_j) - \delta(P'_j,S'_i\setminus P'_j) + \delta(S'_i,P'_j\setminus S'_i)\\
    &\leq \delta(S'_i) - \delta(P'_i) + \delta(S'_i) + \sum_{j\neq i} \delta(P'_j) \\
    &\leq 2 \delta(S'_i) -2\delta(P'_i) + \sum_{j} \delta(P'_j) \leq \sum_{j} \delta(P'_j) - 2w_{min},
\end{align*}
where the last inequality is due to $\delta(P'_i) > 2\delta(S'_i)$ and the minimum non-zero edge weight is $w_{min}$. Thus, the $\ell_1$-norm of the cut values decreases by $2w_{min}$ after each iteration. Since the largest $\ell_1$-norm of the cut values is at most $w_{max}n^2$, the total number of iterations is polynomial in $n$. 

We then show that the partition returned by Algorithm~\ref{alg:uncrossing} satisfies two conditions in the Lemma. We first show that each set $P'_i$ contains at most one terminal.
Note that for every $i = 1, 2,\dots,m''-1$, the set $P'_i$ is a subset of $S'_i \in \calS$. By Lemma~\ref{lem:set_cover}, we have $|P'_i\cap T| \leq |S'_i \cap T| \leq 1$ for all $i = 1, 2,\dots,m''-1$. By Claim~\ref{norm:claim:lower_bound_on_frequency}, every vertex $u\in V$ is covered by at least $\log_2 n$ sets in $\calS$.  By Corollary~\ref{col:number of cover sets}, the cover $\calS$ contains at most $6 k\log_2 n$ sets. Thus, a random set in $\calS$ covers $u$ with probability at least $1/6k$. For each vertex $u \in V$, the probability that $u$ is not covered by any set in $\calS'$ is at most
\begin{equation}\label{eqn:prob_uncovered}
\pr\{u \not \in \cup_{i=1}^{m''-1} S'_i\} \leq \left(1-\frac{1}{6k}\right)^{12k\ln k} \leq \frac{1}{k^2}.
\end{equation}
By the union bound over all terminals, all terminals are covered by $\calS'$ with probability at least $1-1/k$. Thus, the set $P'_{m''}$ contains no terminal with probability at least $1-1/k$.

We now bound the $\ell_1$-norm of the cut values of sets $P'_1,P'_2,\dots,P'_{m''}$.
The first two steps in Algorithm~\ref{alg:uncrossing} can be implemented equivalently by sorting sets in $\calS$ in a random order and picking the first $m''-1 = 12k\ln k$ sets as $\calS'$. Let $S_1,S_2,\dots,S_m$ be the sets in $\calS$ in a random order. Let $\tilde{P}_i = S'_i \setminus \cup_{j<i} S'_j$ for $i=1,2,\dots,m$.  
Then, for any $i =1,2,\dots,m''-1$, the set $\tilde{P}_i$ corresponds to the set $P'_i$ before running the while loop in Algorithm~\ref{alg:uncrossing}.

We first bound the expected $\ell_1$-norm of the cut values of sets $\tilde{P}_1,\tilde{P}_2,\dots,\tilde{P}_{m''-1}$. Note that we have
\[
\sum_{i=1}^{m''-1} \delta(\tilde{P}_i) \leq \sum_{i=1}^{m} \delta(\tilde{P}_i).
\]
We assign each cut edges $(u,v) \in \partial \tilde{P}_i$ into the following two types: (1) edge $(u,v)$ is cut by a set $\tilde{P}_j$ for $j < i$; (2) edge $(u,v)$ is first cut by the set $\tilde{P}_i$. Let $E_i$ be the set of cut edges that first cut by the set $\tilde{P}_i$. Let $w(E_i) = \sum_{e \in E_i} w(e)$ be the total weight of edges in $E_i$.
Each cut edge is counted twice in $\sum_{i=1}^m \delta(\tilde{P}_i)$, while each cut edge is counted exactly once in $\sum_{i=1}^m w(E_i)$. Thus, we have
\[
\sum_{i=1}^{m} \delta(\tilde{P}_i) = 2\sum_{i=1}^{m} w(E_i).  
\]
Note that $E_i \subseteq \partial S_i$ is a subset of edges cut by $S_i$. 
Each edge $(u,v) \in \partial S_i$ is a cut edge in $E_i$ after uncrossing if and only if $S_i$ is the first set among all sets that contain node $u$ or node $v$ in the uncrossing sequence. 
Suppose $S_i$ only contains node $u$. Then, the probability that $(u,v)$ is contained in $E_i$ is at most the probability that $S_i$ is the first set among all the sets that contain node $u$ in the uncrossing sequence. If a set in $\calS$ that contains node $v$ is before set $S_i$, then this edge $(u,v)$ is not count in $E_i$. By Claim~\ref{norm:claim:lower_bound_on_frequency}, we have
\[
\pr\{(u,v) \in E_i\} \leq \pr\{\text{$S_i$ is the first set that contains $u$}\} \leq \frac{1}{\log_2 n}.
\]
Therefore, we have the expected $\ell_1$-norm of the cut values of sets $\tilde{P}_1,\tilde{P}_2,\dots,\tilde{P}_{m}$ is at most
\begin{align*}
\E\left[\sum_{i=1}^{m} \delta(\tilde{P}_i)\right] &= 2\sum_{i=1}^{m} \E [w(E_i)] = 2\sum_{i=1}^{m} \sum_{e \in \partial S_i} w(e) \cdot \pr\{e \in E_i\} \\
&\leq \frac{2}{\log_2 n} \sum_{i=1}^{m} \delta(S_i) \leq k^{1-\nicefrac{1}{p}} \cdot O(\alpha)  \cdot \opt,
\end{align*}
where the last inequality is from Lemma~\ref{lem:set_cover}. At every iteration of the while loop, the $\ell_1$-norm of the cut values of sets $P'_1,P'_2,\cdots,P'_{m''-1}$ only decreases. Thus, we have 
\[
\E\left[\sum_{i=1}^{m''-1} \delta(P'_i)\right] \leq \E\left[\sum_{i=1}^{m''-1} \delta(\tilde{P}_i)\right] \leq k^{1-\nicefrac{1}{p}} \cdot O(\alpha)  \cdot \opt.
\]
Thus, the expected $\ell_1$-norm of the cut values of sets $P'_1,P'_2,\dots,P'_{m''}$ is 
\[
\E\left[\sum_{i=1}^{m''} \delta(P'_i)\right] \leq 2\cdot \E\left[\sum_{i=1}^{m''-1} \delta(P'_i)\right] \leq k^{1-\nicefrac{1}{p}} \cdot O(\alpha)  \cdot \opt.
\]

To bound the $\ell_p$-norm of edge boundaries, we then bound the edge boundary of the last set $P'_{m''}$. 
We only consider the subsampling process in the uncrossing procedure. We sample $O(k \log k)$ sets from the cover $\calS$ uniformly at random. Consider every edge $(u,v)$ in the boundary of sets in cover $\calS$. If this edge $(u,v)$ is a cut edge crossing $P'_{m''}$ and $v \in P'_{m''}$, then one of the sets $S_i \in \calS$ that contains node $u$ is sampled and node $v$ is not covered by sampled sets. Each set $S_i \in \calS$ is sampled with probability $O(\log k/\log n)$.
Suppose the set $S_i\in \calS$ cuts this edge $(u,v)$ and contains node $u$.
Similar to Equation~(\ref{eqn:prob_uncovered}), the probability that node $v \in P'_{m''}$ conditioned on $S_i\in\calS'$ is at most $2/k^2$. Thus, we have 

\begin{align*}
\E[\delta(P'_{m''})] &= \E\left[w\left\{(u,v) \in \bigcup_{i=1}^{m''-1} \partial S'_i: u \not\in P'_{m''} \text{ and } v\in P'_{m''}\right\}\right] \\
&\leq \sum_{i=1}^{m}\sum_{(u,v) \in \partial S_i} w(u,v)\cdot \pr\{ u \in S_i, S_i \in \calS', v\in P'_{m''} \}
\\&
\leq  O\left(\frac{1}{k^2}\cdot\frac{\log k}{\log n}\right)\cdot \sum_{i=1}^{m} \delta(S_i) \leq  O(\alpha)  \cdot \opt,
\end{align*}
where the last inequality is due to condition 4 in Lemma~\ref{lem:set_cover}.

After the while loop, we have $\delta(P'_i) \leq 2\delta(S'_i)$ for all $i =1,2,\dots, m''-1$.  
Since $\E[\delta(P'_{m''})] \leq  O(\alpha)  \cdot \opt$, by Markov's Inequality, we have with probability at least $7/8$ that
$
\delta(P'_{m''}) \leq O(\alpha)  \cdot \opt
$. Since we subsample a fraction $O(\log k/\log n)$ of sets in the cover $S$ uniformly at random, we have 
\begin{align*}
\E\left[\sum_{i=1}^{m''-1} \delta(S'_i)^p\right] \leq O\left(\frac{\log k}{\log n}\right) \sum_{i=1}^{m} \delta(S_i)^p.
\end{align*}
When $\delta(P'_{m''}) \leq O(\alpha)  \cdot \opt$, we have
\begin{align*}
\E\left[\sum_{i=1}^{m''} \delta(P'_i)^p\right] &\leq 2^p\cdot\E\left[\sum_{i=1}^{m''-1} \delta(S'_i)^p\right] + \E\delta(P'_{m''})^p \\
&\leq 2^p \cdot O\left(\frac{\log k}{\log n}\right) \sum_{i=1}^{m} \delta(S_i)^p + O(\alpha^p)  \cdot \opt^p \leq  O(\log k \cdot \alpha^p) \cdot \opt^p,
\end{align*}
where the third inequality is from the condition 3 in Lemma~\ref{lem:set_cover}.
Therefore, we have the conditions 2 and 3 in this lemma hold in expectation with probability at least $7/8$. 
By Markov's Inequality, we have the conditions 2 and 3 in the lemma hold simultaneously with probability at least $3/4$. Since the condition 1 hold with probability at least $1-1/k$, we have all conditions hold with probability at least $3/4-1/k$.
\end{proof}

Next, we analyze the aggregation procedure, which merges these sets to get a $k$ partition of the graph. 

\begin{proof}[Proof of Theorem~\ref{thm:Lp-multiway-cut}]
By Lemma~\ref{lem:uncrossing}, the partition $P'_1,P'_2,\dots, P'_{m'}$ returned by the uncrossing procedure (Algorithm~\ref{alg:uncrossing}) satisfies the following three conditions with probability at least $3/4-1/k$:
\begin{enumerate}
    \item $|P'_i \cap T| \leq 1$ for all $i \in [m']$,
    \item $\sum\limits_{i=1}^{m'}\delta(P'_i)^p \leq  O(\log k\cdot \alpha^p) \cdot \opt^p $,
    \item $\sum\limits_{i=1}^{m'}\delta(P'_i) \leq  O(k^{1-\nicefrac{1}{p}} \cdot \alpha)  \cdot \opt$.
\end{enumerate}

We now assume the partition $\calP'=\{P'_1,P'_2,\dots,P'_{m'}\}$ given by the uncrossing procedure satisfies these three conditions. Then, we use the aggregation procedure as shown in Algorithm~\ref{alg:aggregating} on this partition $\calP'=\{P'_1,P'_2,\dots,P'_{m'}\}$ to get a $k$-partition $\calP=\{P_1,P_2,\cdots,P_k\}$. Since each part $P_i$ has exactly one set $P'_i$ containing one terminal, we have $|P_i \cup T| = 1$ for all $i\in [k]$. 

We now bound the $\ell_p$-norm of the cut values.  
Let $Q'_i = \bigcup_{j > k, Q_j \in \calQ_i} Q_j$ be the union of sets in bucket $\calQ_i$ excluding the set with the largest cut in that bucket. Thus, we have each part $P_i = Q'_i \cup Q_i \cup P'_i$ for all $i \in [k]$. By the triangle inequality, we have
\[
\left(\sum_{i=1}^k \delta(P_i)^p\right)^{1/p} \leq \left(\sum_{i=1}^k \delta(Q'_i)^p\right)^{1/p} + \left(\sum_{i=1}^k \delta(Q_i)^p\right)^{1/p} + \left(\sum_{i=1}^k \delta(P'_i)^p\right)^{1/p}.
\]
By Lemma~\ref{lem:uncrossing}, the $\ell_p$-norm of the cut values of sets $P'_1,P'_2,\dots,P'_{k}$ is 
\[
\left(\sum_{i=1}^k \delta(P'_i)^p\right)^{1/p} \leq O(\log^{1/p} k \cdot \alpha) \cdot \opt.
\]
Similarly, we have the $\ell_p$-norm of the cut values of sets $Q_1,Q_2,\dots,Q_{k}$ is
\[
\left(\sum_{i=1}^k \delta(Q_i)^p\right)^{1/p} \leq O(\log^{1/p} k \cdot \alpha) \cdot \opt.
\]
We then bound the $\ell_p$-norm of the cut values of sets $Q'_1,Q'_2,\dots,Q'_{k}$. We first bound the cut value of each set $Q'_i$. Since $Q_i$ are sorted by the cut value in descending order,  we have
\[
\delta(Q'_i) \leq \sum_{j > k, Q_j \in \calQ_i} \delta(Q_j) \leq \sum_{Q_j \in \calQ_k} \delta(Q_j) \leq \frac{1}{k} \sum_{Q_j \in \calQ} \delta(Q_j), 
\]
where the second inequality is due to $\delta(Q_{i + zk}) \leq \delta(Q_{zk})$ for $z \geq 1$ and the third inequality is because $\calQ_k$ contains the smallest cut set for every $k$ consecutive sets. By Lemma~\ref{lem:uncrossing}, we have 
\[
\delta(Q'_i) \leq \frac{1}{k} \cdot  \sum_{Q_j \in \calQ} \delta(Q_j) \leq O(k^{-1/p}\cdot \alpha) \cdot \opt.
\]
Therefore, we have $\ell_p$-norm of the cut values of sets $Q'_1,Q'_2,\dots,Q'_{k}$ is at most
\[
\left(\sum_{i=1}^k \delta(Q'_i)^p\right)^{1/p} \leq \left(k \cdot O(k^{-1}\cdot \alpha^p) \cdot \opt^p \right)^{1/p} = O(\alpha) \cdot \opt. 
\]
Combining three parts, we get the conclusion.
\end{proof}

By Theorem~\ref{thm:Lp-multiway-cut}, given a graph with $n$ vertices and $k$ terminals, our algorithm finds an\\ $O(\log^{\nicefrac{1}{2}} n \log^{\nicefrac{1}{2}+\nicefrac{1}{p}} k)$ approximation for the $\ell_p$-Norm Multiway Cut with probability at least $3/4-1/k$. We can repeat this algorithm $O(\log 1/\varepsilon)$ times to find an $O(\log^{\nicefrac{1}{2}} n \log^{\nicefrac{1}{2}+\nicefrac{1}{p}} k)$ approximation for the $\ell_p$-Norm Multiway Cut with probability at least $1-\varepsilon$, which proves Theorem~\ref{thm:lp}. 

\section{Norm Multiway Cut}\label{sec:Norm Multiway Cut}
In this section, we provide approximation algorithms for the Norm Multiway Cut under an arbitrary monotonic norm, $\norm{\cdot}$. We prove Theorem~\ref{thm:minimization-oracle} in Sections~\ref{apx:covering} and \ref{apx:aggregation}, and Theorem~\ref{thm:ordering-oracle} in Section~\ref{apx:improve}.

\newtheorem*{thm:minimization-oracle}{Theorem \ref{thm:minimization-oracle}}
\begin{thm:minimization-oracle}
    There exists a polynomial-time  algorithm that for every monotonic norm with a minimization oracle, given a graph with $n$ vertices, $k$ terminals, finds an $O(\log^{\nicefrac{1}{2}} n \log^{\nicefrac{7}{2}} k)$ approximation for the Norm Multiway Cut with high probability.
\end{thm:minimization-oracle}

\newtheorem*{thm:ordering-oracle}{Theorem \ref{thm:ordering-oracle}}
\begin{thm:ordering-oracle}
    There exists a polynomial-time algorithm that for every monotonic norm with an ordering oracle,  given a graph with $n$ vertices, $k$ terminals, finds an $O(\log^{\nicefrac{1}{2}} n \log^{\nicefrac{5}{2}} k)$ approximation for the Norm Multiway Cut with high probability.
\end{thm:ordering-oracle}

To solve the general Norm Multiway Cut problem, we slightly modify the algorithm for the $\ell_p$-Norm Multiway Cut problem. Our algorithm for the Norm Multiway Cut problem uses the same framework with three parts: covering, uncrossing, and aggregation procedures. 
We first describe our algorithm with a minimization oracle in Sections~\ref{apx:covering} and \ref{apx:aggregation}. We then show an improved algorithm with an ordering oracle in Section~\ref{apx:improve}. 
We start with describing the covering procedure.

\subsection{Covering Procedure}\label{apx:covering}

We first define some notations used in our covering procedure. We have a minimization oracle for the monotonic norm in the Norm Multiway Cut. Let $A_i$ be the set of $i$ coordinates that minimizes the norm of the indicator vector.
For $0 \leq i\leq \floor{\log_2 k}-1$, let $I_i = A_{2^i}$. When $k > 2^{\floor{\log_2 k}}$, let $I_{\floor{\log_2 k}}=\argmin_{I\subseteq [k], \vert I\vert = k-2^{\floor{\log_2 k}}}\norm{\one_{I}}$. These sets $I_i$ can be efficiently computed by the minimization oracle. These sets $\{I_{0},\dots,I_{\floor{\log_2 k}}\}$ do not necessarily form a partition of $\{1,2,\cdots,k\}$. Let $\calP^*=\{P_1^*,\dots,P_k^*\}$ be an optimal solution to a Norm Multiway Cut instance. Let $\opt=\norm{(\delta(P^*_1),\cdots,\delta(P^*_k))}$ denote the optimal objective value. We define $r_i=\nicefrac{\opt}{\norm{\one_{I_i}}}$ for $0\leq i\leq\floor{\log_2 k}$. Note that $\opt$ can be guessed within a factor of two via a binary search. Thus, for the rest of the section, we assume that $\opt$ is known.



The covering procedure is similar to the algorithm we used for minimizing the $\ell_p$-Norm Multiway Cut. 
Unlike the $\ell_p$-norm, the general monotonic norm is not necessarily permutation invariant, which means the norm of a vector depends on the ordering of this vector. 
Thus, for each terminal $t_i$, we compute a subset of vertices $C_i\subseteq V$ isolating $t_i$ from other terminals with minimum cut value; and ``remove'' them from the graph.  More formally, let $C_i=\argmin_{C\subseteq V, C\cap T = \{t_i\}}\delta(C)$ for each $i\in [k]$, and define $\calC = \{C_1,\dots, C_k\}$. Then, $\norm{(\delta(C_1),\dots,\delta(C_k))} \leq \opt$ since the norm is monotonic and $\delta(C_i)\leq\delta(P^*_i)$ for all $i\in [k]$. Thus, we can focus on covering the remaining graph $V\setminus \bigcup_{i=1}^{k}C_i$, which does not contain any terminal.
 
Next we produce a collection of sets $\calS=\{S_1,\dots,S_m\}$ which covers $V\setminus \bigcup_{i=1}^{k}C_i$. Each set $S_i\in\calS$ contains at most one terminal, has a large measure, and induces a small cut. As in the case of $\ell_p$-Norm Multiway Cut, the covering algorithm relies on the multiplicative weights update method. 
It initializes the measure of each vertex to one. At each iteration $t$, the algorithm computes a set $S_t$ using the Unbalanced Terminal Cut (UTC) algorithm in Theorem~\ref{thm:covering:UTC} as a subroutine, and reduces the measure of the vertices covered by it by half. 
The algorithm terminates when the total measure of vertices is less than $\nicefrac{1}{n}$. 
To compute $S_t$, it runs the UTC algorithm $O(\log k)$ many times with different measure parameters and outputs the one inducing a cut smaller than a certain threshold. 
More specifically, at time $i$ the covering algorithm calls UTC with measure parameter $\rho_i = \nicefrac{1}{2\log_2 k\cdot \vert I_i \vert}$ and checks if its output, $U_i$, satisfies $\delta(U_i)\leq \alpha\cdot r_i$ where $\alpha=O\left(\sqrt{\log n\log k}\right)$ (see Theorem~\ref{thm:covering:UTC}). 
We show this algorithm in Algorithm~\ref{alg:arbitrary_norm_covering}.
\begin{algorithm}
   \caption{Covering Procedure for Norm Multiway Cut}
   \label{alg:arbitrary_norm_covering}
\begin{algorithmic}
   \STATE Let $C_i$ be a minimum cut isolating $t_i$ from remaining terminals for $i\in [k]$.
   \STATE Set $t=1$, and $\mu_1(v)=1$ for all $v\in V$. Let $\calS = \emptyset$. \\
   \WHILE{$\sum\limits_{v\in V}\mu_t(v)\geq \frac{1}{n}$}
   \FOR{$i = 0,\dots, \floor{\log_2 k}$}
   \STATE Let $U_i\subseteq V$ be the solution for Unbalanced Terminal Cut instance $\langle G,w,\mu_t,\frac{1}{2\log_2 k\cdot \vert I_i  \vert},T \rangle$.
   \IF{$\delta(U_i)\leq \alpha\cdot r_i $}
   \STATE $S_t=U_i$
   \STATE \textbf{break}
   \ENDIF
   \ENDFOR
   \STATE Let $\calS = \calS \cup \{S_t\}$.
   \FOR{$v\in V$}
   \STATE \text{Set }
$
\mu_{t+1}(v) = 
     \begin{cases}
       \nicefrac{\mu_t(v)}{2} &\quad\text{if } v\in S_t,\\
       \mu_t(v) &\quad\text{if } v \not\in S_t.
     \end{cases}
$
   \ENDFOR
   \STATE Set $t = t+1$
   \ENDWHILE
   \RETURN $\calS\cup\{C_1,\dots,C_k\}$.
\end{algorithmic}
\end{algorithm}

 Next we prove that Algorithm~\ref{alg:arbitrary_norm_covering} terminates, i.e., it always finds a set $S_t$ for each iteration $t$. For this we show the existence of a set in an optimal solution with a large measure and a small cut value.
 \begin{lemma}\label{arbit:lem:optimal_set_exists}
Let $\calP^* = \{P_1^*,\dots,P_k^*\}$ be an optimal solution to a Norm Multiway Cut instance and let $\opt$ denote the norm of $\calP^*$. Let $\mu:V\rightarrow \mathbb{R}_{\geq 0}$ be any measure on vertices. There exist indices $0\leq i\leq \floor{\log_2 k}$ and $j\in[k]$ such that the following two conditions hold:
\begin{enumerate}
    \item $\mu(P_j^*)\geq \frac{\mu(V)}{2\log_2 k\cdot \vert I_i \vert}$
    \item $\delta(P_j^*)\leq r_i$
\end{enumerate}
\end{lemma}
\begin{proof}
Let $\pi$ be a permutation on $k$ coordinates so that the sets in an optimal solution, $\calP^*$, are ordered in a decreasing order in terms of their cut values. In other words, $\delta(P^*_{\pi^{-1}(i)})\geq \delta(P^*_{\pi^{-1}(i+1)})$ for all $i$. 
Without loss of generality, we assume that $\pi$ is identity, which means, $\delta(P^*_1)\geq\dots\geq \delta(P^*_k)$.

We partition all coordinates $\{1,2,\cdots,k\}$ into $\floor{\log_2 k}+2$ buckets $B_{-1},B_0,\dots,B_{\floor{\log_2 k}}$ according to the cut value $\delta(P^*_i)$. Define $B_{-1} = \{1\}$, and $B_i = \{2^i+1,\cdots, \min(k,2^{i+1})\}$ for $0\leq i\leq \floor{\log_2 k}$. Note that the size of buckets are exponentially increasing, $\vert B_{-1} \vert = 1$, $\vert B_{\floor{\log_2 k}}\vert = k - 2^{\floor{\log_2 k}}$, and $\vert B_i \vert = 2^i$ for $-1 < i < \floor{\log_2 k}$. If $k$ is a power of two, $B_{\floor{\log_2 k}}$ is an empty set. 
For any buckets $B_i$ and $B_{i^{'}}$ such that $i^{'}<i$, we have for any indices $j\in B_i$ and $j^{'}\in B_{i^{'}}$ we have $\delta(P^*_j)\leq \delta(P^*_{j^{'}})$ since $\calP^*$ is sorted in a decreasing order. 
For any $i > -1$, let $\tilde{B}_i = \bigcup_{l=-1}^{i-1}B_l$ be the union of all buckets $B_l$ with $l < i$.
From the monotonicity of the norm, it follows that for any $-1 < i < \floor{\log_2 k}$ and any index $j\in B_i$, $\norm{\delta(P^*_j)\cdot \one_{\tilde{B}_i}}\leq \norm{\calP^*}=\opt$.
The bucket $B_{-1}$ satisfies $ \norm{\delta(P^*_1)\cdot \one_{{B_{-1}}}} \leq \opt$. For any index $j\in B_{\floor{\log_2 k}}$, we have $\norm{\delta(P^*_j)\cdot \one_{S}}\leq \opt$ for any set $S\subseteq B_{\floor{\log_2 k}-1}$ with $\vert S \vert = \vert B_{\floor{\log_2 k}} \vert = k-2^{\floor{\log_2 k}}$. 

We now show the existence of a set $P^*_i$ satisfying two conditions. Recall that the set $I_i$ denote the set of $2^i$ coordinates that minimizes the norm over all indicator vectors of $2^i$ coordinates. Thus, it follows that (1) $\norm{\delta(P^*_1)\cdot\one_{I_0}}\leq \norm{\delta(P^*_1)\cdot \one_{{B_{-1}}}}$; (2) $\norm{\delta(P^*_j)\cdot \one_{I_{\floor{\log_2 k}}}}\leq\norm{\delta(P^*_j)\cdot \one_{S}}$ for any index $j\in B_{\floor{\log_2 k}}$ and any set $S\subseteq B_{\floor{\log_2 k}-1}$ with $\vert S \vert = \vert B_{\floor{\log_2 k}} \vert = k-2^{\floor{\log_2 k}}$; and (3) since $|\tilde{B}_i| = 2^i$ for $-1< i < \floor{\log_2 k}$, $\norm{\delta(P^*_j)\cdot \one_{I_i}} \leq \norm{\delta(P^*_j)\cdot \one_{\tilde{B}_i}}$ for an index $j\in B_i$. 
This implies that for any bucket $B_i$ and for any index $j\in B_i$, the cut value  $\delta(P^*_j)\leq \frac{\opt}{\norm{1_{I_i}}} = r_i$. This means all sets $P^*_i$ satisfies Condition $2$.

It remains to find a set $P^*_i$ satisfying Condition $1$. Recall that $\{B_{-1},\dots B_{\floor{\log_2 k}}\}$ is a partitioning of coordinates $[k]$ into at most $\floor{\log_2 k}+2$  buckets. Thus, there exists a bucket $B_{i^{*}}$ such that $\sum_{j\in B_{i^{*}}}\mu(P^*_j)\geq \frac{\mu(V)}{\floor{\log_2 k}+2}\geq \frac{\mu(V)}{2\log_2 k}$. By the definition of sets $\{I_i\}$, we have $\vert B_{-1} \vert = \vert I_0 \vert$, and $\vert B_{i} \vert = \vert I_i \vert$ for all $i\geq 0$. This implies that there exists an index $j\in B_{i^{*}}$ so that $\mu(P^*_{j})\geq \frac{\mu(V)}{2\log_2 k\cdot \vert B_{i^{*}} \vert}=\frac{\mu(V)}{2\log_2 k\cdot \vert I_{i^{*}} \vert}$, i.e. $P^*_j$ satisfies Condition 1.
\end{proof}

By Lemma~\ref{arbit:lem:optimal_set_exists}, we know that Algorithm~\ref{alg:arbitrary_norm_covering} terminates. 
Let $\mu_t$ be the measure on $V$ at iteration $t$ of Algorithm~\ref{alg:arbitrary_norm_covering}. Let $i$ and $j$ be indices so that $P^*_j$ satisfies the conditions in Lemma~\ref{arbit:lem:optimal_set_exists} with respect to $\mu_t$. 
Thus, the set $U_i$ returned by the $i$th call to the UTC algorithm satisfies $\delta(U_i)\leq\alpha\cdot\delta(P^*_j)\leq \alpha \cdot r_i$, which means Algorithm~\ref{alg:arbitrary_norm_covering} always finds the required set $S_t$.


We group the sets in the cover $\calS$ according to the measure constraint used in the UTC algorithm. 
Let $G_i\subseteq \calS$ be a collection of sets output by Algorithm~\ref{alg:arbitrary_norm_covering} due to the $i$-th call to the UTC algorithm for $0\leq i\leq \floor{\log_2 k}$. Each group $G_i$ corresponds to a bucket $B_i$.
Applying Claim~\ref{clm:covering:total_measure_bound} on $G_i$ we get the following bound on its size.


\begin{corollary}\label{arbit:corol:bound_on_sets}
$\vert G_i\vert \leq O(\vert I_i \vert \log k \log n)$ for $0\leq i\leq \floor{\log_2 k}$.
\end{corollary}

Since $|I_i| = 2^i$, the cover $\calS$ contains at most $6k\log_2 k\log_2 n = O(k\log k\log n)$ sets.

\subsection{Uncrossing and Aggregation Procedures}\label{apx:aggregation}
In this section, we show procedures that transform the cover $\calS\cup \calC$ produced by Algorithm~\ref{alg:arbitrary_norm_covering} to a $k$ partition of the graph $\calP = \{P_1,P_2,\cdots, P_k\}$. We show that this partition is an $O(\log^{\nicefrac{1}{2}} n \log^{\nicefrac{7}{2}} k)$ approximation to the Norm Multiway Cut.

We first use an uncrossing procedure similar to Algorithm~\ref{alg:uncrossing} to generate a partition of the graph. 
Recall that Algorithm~\ref{alg:arbitrary_norm_covering} returns $\calC\cup \calS$ where $\calC=\{C_1,\dots,C_k\}$ is the collection of minimum isolating cuts for terminals, and $\calS=\{S_1,\dots,S_m\}$ is the collection of sets produced by the multiplicative weights update. For the Norm Multiway Cut, we keep all sets in $\calC$ and only sample sets in $\calS$. Then, we create a sequence of sets with all sets in $\calC$ as the first $k$ sets, which are followed by sampled sets from $\calS$ sorted in random order. Next, we run the same uncrossing through this sequence. We show this uncrossing procedure in Algorithm~\ref{arbit:alg:uncrossing}.

\begin{algorithm}
    \caption{Uncrossing Procedure for Norm Multiway Cut}
    \label{arbit:alg:uncrossing}
\begin{algorithmic}
    \STATE \textbf{Input} $\calC\cup \calS$.
    \STATE Sample $m''-1=9k\ln^2 k$ sets $\calS' = (S'_1,S'_2,\dots,S'_{m''-1})$ from $\calS$ uniformly at random.
    \STATE Let $Z_1,Z_2,\dots,Z_{m''+k-1}$ be an ordering of sets in $\calC\cup\calS^{'}$ such that $C_1,\dots,C_k$ occupy the first $k$ positions, respectively.
    \STATE Set $P'_i = Z_i \setminus \bigcup\limits_{j<i} Z_j$ for all $i$.
    \WHILE{there exists a set $P'_i$ such that $\delta(P'_i) > 2\delta(Z_i)$}
    \STATE Set $P'_i = Z_i$, and set $P'_j = P'_j \setminus Z_i$ for all $j\neq i$.
    \ENDWHILE
    \STATE Let $P'_{m''+k} = V \setminus \bigcup_{i=1}^{m''+k-1} P'_i$.
    \RETURN all non-empty sets $P'_i$ for $i \in [m''+k]$.
\end{algorithmic}
\end{algorithm}

We first provide some useful properties of the partition given by the uncrossing procedure. Let $\calP' = \{P'_1,\dots, P'_{m'}\}$ be the partition of the graph returned by Algorithm~\ref{arbit:alg:uncrossing}. For simplicity, we assume that $\calP'$ includes those empty sets which are removed in the last step of Algorithm~\ref{arbit:alg:uncrossing}. 
Observe that $P'_i\subseteq Z_i$ and $\delta(P'_i)\leq 2\delta(Z_i)$ for all $1 \leq i \leq m''+k-1$. We get the following corollary immediately.
\begin{corollary}\label{arbit:corol:bound_min_isolating}
$\norm{(\delta(P'_1),\dots,\delta(P'_k))}\leq 2\cdot \norm{(\delta(C_1),\dots,\delta(C_k))}\leq 2\opt$.
\end{corollary}

It remains to give an upper bound on the cut value of the last set $P'_{m'}$. Recall that $G_i\subseteq \calS$ is a collection of sets output by the covering procedure due to the $i$-th call to the UTC algorithm, which corresponds to bucket $B_i$. Let $G^{'}_i\subseteq \calP'$ be a collection of sets produced by Algorithm~\ref{arbit:alg:uncrossing} due to $G_i$ for $0\leq i\leq \floor{\log_2 k}$. We start with the following bound on the size of $G^{'}_i$. Note that the cover $\calS$ has at most $6k\log_2 k\log_2 n$ sets. Since we uniformly sampled $9k\ln^2 k$ sets from $\calS$, each set $S_i \in \calS$ is sampled with probability at most $\log_2 k/\log_2 n$. By Corollary~\ref{arbit:corol:bound_on_sets}, we have
\begin{corollary}\label{arbit:corol:expected_bound_on_partition_size}
$\E[\vert G^{'}_i \vert]\leq O\left(\vert I_i\vert \log^2 k\right)$. 
\end{corollary}
Now we give an upper bound on the probability that for every vertex $v\in V$, $v$ is contained in $P'_{m'}$.
\begin{claim}\label{arbit:claim:probability_of_being_in_S}
$\Pr(v\in P'_{m'})\leq \frac{1}{k}$ for every vertex $v\in V$.
\end{claim}
\begin{proof}
Let $N_v$ denote the number of sets in $\calS$ containing $v$. From Claim~\ref{norm:claim:lower_bound_on_frequency} it follows that $N_v\geq \log_2 n$. Then, we have
\begin{align*}
    \Pr(v\in P'_{m'}) &\leq \left(1-\frac{\log_2 n}{6k\log_2 k\log_2 n}\right)^{9k \ln^2 k}\\
    &\leq\left(1-\frac{1}{6k\log_2 k}\right)^{6 k \log_2 k \ln k} 
    \leq e^{-\ln k} = \frac{1}{k}.
\end{align*}
\end{proof}

Note that for an edge $(u,v)$, the event $(u,v)\in \delta(P'_{m'})$ occurs only when exactly one of its endpoints belongs to $P'_{m'}$ and this edge $(u,v)$ is in the cut of $\bigcup_{i=1}^{m'-1} \partial P'_i$. Let $r=\max_{0\leq i\leq \floor{\log_2 k}}r_i$. Now we give an upper bound on the expected cut value of $P'_{m'}$, $\E[\delta(P'_{m'})]$.

\begin{claim}\label{arbit:clm:bound_on_S}
$\E[\delta(P'_{m'})]\leq O\left(\log^2 k\right)\cdot\alpha r$.
\end{claim}

\begin{proof}
By Claim~\ref{arbit:claim:probability_of_being_in_S}, each vertex is contained in $P'_{m'}$ with probability at most $1/k$. Thus, we have
\begin{align*}
    \E[\delta(P'_{m'})]&\leq \frac{2}{k}\cdot\E\left[\sum_{i=1}^{m'-1}\delta(P'_i)\right]\leq \frac{2}{k}\cdot \alpha r\cdot O(k\log^2 k) = O\left(\log^2 k\right)\cdot \alpha r,
\end{align*}
where the second inequality follows from $\delta(P'_i) \leq \alpha r$ for all $1\leq i \leq m'-1$.
\end{proof}


We now present a modified aggregation procedure that converts the partition $\calP'$ given by Algorithm~\ref{arbit:alg:uncrossing} to a $k$ partition of the graph $\calP$. Recall that Algorithm~\ref{arbit:alg:uncrossing} returns a partition $\calP'$ so that the first $k$ sets of $\calP'$, $\{P'_1,\dots,P'_k\}$, correspond to $k$ minimum isolating cuts $\calC=\{C_1,\dots,C_k\}$. Furthermore, recall that $G_i\subseteq \calS$ is a collection of sets output by Algorithm~\ref{alg:arbitrary_norm_covering} due to the $i$-th call to the UTC algorithm, which corresponds to bucket $B_i$. Let $G^{'}_i\subseteq \calP'$ be a collection of sets produced by Algorithm~\ref{arbit:alg:uncrossing} due to $G_i$ for $0\leq i\leq \floor{\log_2 k}$.

Firstly, the aggregation procedure partitions sets in $G^{'}_i$ into $\vert I_i \vert$ subgroups with equal size, denoted by $\{R_{j}^{i}\}_{j\in I_i}$. Each subgroup $R_j^{i}$ corresponds to a coordinate $j\in I_i$. Let $i^*=\argmax_{0\leq i\leq \floor{\log_2 k}} r_i$ and let $j_s$ denote the smallest coordinate in $I_{i^*}$, i.e., $j_s=\min_{j\in I_{i^*}}j$. Then for each $j \in [k]$, it sets
\[   
P_j = 
     \begin{cases}
       P'_j\cup\bigcup\limits_{j\in I_i}\bigcup\limits_{P'\in R^i_{j}}P' &\quad\text{if } j\neq j_s,\\
       P'_j\cup P'_{m'} \cup\bigcup\limits_{j\in I_i}\bigcup\limits_{P'\in R^i_{j}}P' &\quad\text{if } j= j_s.
     \end{cases}
\]
We give a pseudo-code for this algorithm in Algorithm~\ref{arbit:alg:merging}.
\begin{algorithm}
    \caption{Aggregation Procedure for Norm Multiway Cut}
    \label{arbit:alg:merging}
\begin{algorithmic}
    \STATE \textbf{input} $\calP'$.
    \STATE Partition $G^{'}_i$ into $\vert I_i \vert$ equal-size sets, $\{R_{j}^{i}\}_{j\in I_i}$, so that $R_j^{i}$ corresponds to a coordinate $j\in I_i$.
    \STATE Let $i^*=\argmax\limits_{0\leq i\leq \floor{\log_2 k}}r_i$ and $j_s=\min\limits_{j\in I_{i^*}}j$.
    \FOR{$j=1,\cdots,k$}
    \STATE 
\[   
P_j = 
     \begin{cases}
       P'_j\cup\bigcup\limits_{j\in I_i}\bigcup\limits_{P'\in R^i_{j}}P' &\quad\text{if } j\neq j_s,\\
       P'_j\cup P'_{m'} \cup\bigcup\limits_{j\in I_i}\bigcup\limits_{P'\in R^i_{j}}P' &\quad\text{if } j= j_s.
     \end{cases}
\]
    \ENDFOR
    \RETURN $\calP=\{P_1,\dots,P_k\}$.
\end{algorithmic}
\end{algorithm}

Now we prove that the expected norm of partition $\calP$ is small, i.e., $\E[\norm{(\delta(P_1),\cdots,\delta(P_k))}]\leq O(\log^3 k)\cdot \alpha\cdot\opt$ where $\alpha=O\left(\sqrt{\log n\log k}\right)$.

\begin{proof}[Proof of Theorem~\ref{thm:minimization-oracle}]
Note that each set $P_j$ in the partition $\calP$ contains at most three types of vertices. The first type is the vertices in the minimum isolating cut $P'_j$. The second type is the vertices in subgroup $R^i_j$. The third type only exists for $P_{j_s}$, which is the vertices in $P'_{m'}$. 
For every $i \leq 0\leq i\leq \floor{\log_2 k}$, we define vector $v^i$ to denote the cut values of vertices in subgroups $R^i_j$. For every coordinate $j \in I_i$, we have $v^i_j = \delta(\bigcup_{P'\in R^i_j} P')$. For $j \not \in I_i$, we have $v^i_j = 0$.
Let $\tilde{v} = \delta(P'_{m'})\cdot\one_{\{j_s\}}$ be the vector corresponds to the third type of vertices in $P'_{m'}$.
Then, from triangle inequality and linearity of expectation, it follows that
\begin{align}\label{arbit:ineq_triangle}
    \E[\norm{(\delta(P_1),\cdots,\delta(P_k))}]\leq \E[\norm{(\delta(P'_1),\cdots,\delta(P'_k))}]+\sum\limits_{i=0}^{\floor{\log_2 k}}\E[\norm{v^i}]+\E[\norm{\tilde{v}}].
\end{align}
For the last part $P'_{m'}$ in the $\calP'$, by Corollary~\ref{arbit:clm:bound_on_S}, we have $\E[\delta(P'_{m'})]=O(\log^2 k)\cdot\alpha r$ where $r=\max_{0\leq i\leq \floor{\log_2 k}}r_i$. Note that $j_s \in I_{i^*}$. Thus, we have
\begin{align*}
\E[\norm{\tilde{v}}]\leq \E[\delta(P'_{m'})] \norm{\one_{I_{i^*}}}\leq O(\log^2 k)\cdot\alpha \cdot r\cdot\norm{\one_{I_{i^*}}}= O(\log^2 k)\cdot\alpha \cdot \opt.
\end{align*}
where the last equality follows from $r = r_{i^*} = \nicefrac{\opt}{\norm{\one_{I_{i^*}}}}$.

Next we give an upper-bound on $\E[\norm{v^i}]$ for $0\leq i\leq \floor{\log_2 k}$. Note that every set $P'_l \in R^i_j$ is in group $G'_i$, which implies $\delta(P'_l) \leq 2\alpha r_i$ according to the uncrossing procedure. By the definition of vector $v^i$, we have for each coordinate $j\in I_i$
\begin{align*}
    v^i_j= \delta\left(\bigcup\limits_{P'\in R^i_{j}}P'\right)\leq \sum\limits_{P'\in R^i_j}\delta(P')\leq \vert R_j^i \vert\cdot 2\alpha r_i.
\end{align*}
For each coordinate $j \not\in I_i$, $v^i_j = 0$.
The monotonicity of the norm implies
\begin{align}\label{arbit:ineq:bound_M_i}
\E[\norm{v^i}]\leq \E[|R^i_j|]\cdot 2\alpha r_i\cdot  \norm{\one_{I_{i}}}\leq O(\log^2 k)\cdot\alpha\cdot r_i\cdot\norm{\one_{I_{i}}}= O(\log^2 k)\cdot\alpha \cdot \opt.
\end{align}

By Corollary~\ref{arbit:corol:bound_min_isolating}, we have the norm of the minimum isolating cuts is at most $2\opt$. Combining the upper bounds for three parts in Equation~(\ref{arbit:ineq_triangle}), we have
\begin{align*}
    \E[\norm{(\delta(P_1),\cdots,\delta(P_k))}]\leq O(\log^3 k)\cdot\alpha\cdot \opt = O(\log^3 k\cdot \sqrt{\log n \log k})\cdot \opt,
\end{align*}
which finalizes the proof.    
\end{proof}

\subsection{Improved Analysis with Strong Oracle}\label{apx:improve}

In this section, we consider the Norm Multiway Cut under a monotonic norm with an ordering oracle. For any vector $v$, the ordering oracle can provide the ordering of this vector that minimizes the norm.

We slightly modify the algorithm for Norm Multiway Cut with a minimization oracle. 
Let $\calP^*=\{P^*_1,\cdots, P^*_k\}$ be the optimal solution. 
We sort these sets by their cut values in descending order such that $\delta(P^*_1) \geq \delta(P^*_2) \geq \cdots \geq \delta(P^*_k)$. 
We partition all coordinates $\{1,2,\cdots,k\}$ into $\floor{\log_2 k}+1$ buckets $B'_0,\dots,B'_{\floor{\log_2 k}}$ according to the cut value $\delta(P^*_i)$. We define $B'_i = \{2^i,\cdots, \min(k,2^{i+1}-1)\}$ for $0\leq i\leq \floor{\log_2 k}$. The size of $B'_i$ increasing exponentially, $|B'_i| = 2^i$.

For each bucket $B'_i$, let $b_i = \delta(P^*_{2^{i}})$ for $0 \leq i \leq \floor{\log_2 k}$ be the largest cut value in the bucket $B'_i$.  Suppose we know $b_i$ for all $0 \leq i \leq \floor{\log_2 k}$. Then, we use $b_i$ instead of $r_i$ for the measure constraint in the UTC algorithm in the covering procedure (See Algorithm~\ref{alg:arbitrary_norm_covering}). We discuss later how to guess $b_i$ efficiently. After the covering procedure, we run the same uncrossing procedure in Algorithm~\ref{arbit:alg:uncrossing}. Then, we modify the aggregation procedure in Algorithm~\ref{arbit:alg:merging} to generate a $k$ partition. For the minimization oracle, we partition sets in each group $G'_i$ into $|I_i|$ subgroups $R^i_j$ corresponding to coordinates in $j \in I_i$. For the ordering oracle, we assign each subgroup $R^i_j$ to a coordinate $j \in [k]$ according to the following claim.

\begin{claim}\label{clm:ordering}
    Consider a monotonic norm, $\norm{\cdot}$. For any vector $v$, let $x_1 \geq x_2 \geq \cdots \geq x_k$ be the entries of vector $v$ sorted in descending order. For every $0 \leq i \leq \floor{\log_2 k}$, let $b_i = x_{2^{i}}$. There exists an assignment of $b_i$ to $2^i$ coordinates $y^i_1,y^i_2,\dots, y^i_{2^i} \in [k]$ such that
    \[
    \norm{\sum_i b_i \sum_{j=1}^{2^i} \one_{y^i_j}} \leq 3 \norm{v}.
    \]
\end{claim}
\begin{proof}
We create the following three vectors in $\bbR^k$, $u'_1$, $u'_2$, and $u'_3$. Let the vector $u'_1 = b_1 \one_1$ that contains only one non-zero entry $b_1$. Let vectors $u'_2=u'_3$ contain $2^{i-1}$ entries with value $b_i$ for $1\leq i \leq \floor{\log_2 k}$.  By the monotonicity of the norm, we have $\norm{u'_1} \leq \norm{v}$, $\norm{u'_2} \leq \norm{v}$, and $\norm{u'_3} \leq \norm{v}$. Let $u_1,u_2,u_3$ be the ordering of vectors $u'_1$,$u'_2$, and $u'_3$ that minimizes the norm respectively. Then, we have
    \[
    \norm{u_1+u_2+u_3} \leq \norm{u_1} + \norm {u_2}+\norm{u_3} \leq \norm{u'_1} + \norm {u'_2}+\norm{u'_3} \leq 3\norm{v}. 
    \]
Since vectors $u_1$, $u_2$, and $u_3$ contains total $2^i$ entries of $b_i$ for all buckets $i$, thus $u_1+u_2+u_3$ is the required assignment.
\end{proof}

With the ordering oracle, we can efficiently find the assignment shown in Claim~\ref{clm:ordering}. Let $I'_i = \{y^i_1,y^i_2,\dots, y^i_{2^i}\}$ be the multiset of coordinates that $b_i$ is assigned to. Now we show that the partition $\calP=\{P_1,P_2,\cdots, P_k\}$ returned by the above algorithm is a good approximation for the Norm Multiway Cut.


\begin{proof}[Proof of Theorem~\ref{thm:ordering-oracle}]

Since each set $P_i$ contains exactly one minimum isolating cut $P'_i$, this set $P_i$ contains exactly one terminal. We then bound the norm of this solution. 
Similar to Theorem~\ref{thm:minimization-oracle}, we partition all vertices into three types. For every $0\leq i\leq \floor{\log_2 k}$, We let $v^i = b_i \sum_{j \in I'_i} \one_{\{j\}}$ be the vector of cut values corresponding to sets in $G'_i$. Let $\tilde{v} = \delta(P'_{m'})\cdot\one_{\{j_s\}}$ be the vector corresponds to the third type of vertices in $P'_{m'}$.
By triangle inequality and linearity of expectation, we have
\begin{align}\label{arbit:ineq_triangle_new}
    \E[\norm{(\delta(P_1),\cdots,\delta(P_k))}]\leq \E[\norm{(\delta(P'_1),\dots,\delta(P'_k))}]+\E\left\{\left\lVert\sum_{i=0}^{\floor{\log_2 k}} v^i \right\rVert\right\}+\E[\norm{\tilde{v}}].
\end{align}
Since for any $0\leq i \leq \floor{\log_2 k}$ and  any coordinate $j \in I'_i$, we have $\E[v^i_j] \leq O(\log ^2 k) \alpha b_i$. By Claim~\ref{clm:ordering}, we have
\begin{align*}
    \E\left\{\left\lVert\sum_{i=0}^{\floor{\log_2 k}} v^i \right\rVert\right\} \leq O(\log^2 k \cdot \alpha)\left\lVert\sum_i b_i \sum_{j \in I'_i}\one_{\{j\}}\right\rVert \leq O(\log^2 k \cdot \alpha) \opt.
\end{align*}
Since the first and third terms in the right-hand side of Equation~(\ref{arbit:ineq_triangle_new}) is upper bounded by $O(\log^2 k \cdot \alpha) \opt$ in Section~\ref{apx:aggregation}, we have  $\calP$ is an $O(\log^2 k \cdot \alpha)$ approximation for the Norm Multiway Cut. 

Now we specify how to guess values $b_i$ efficiently. First, we guess the optimal value $\opt$. Let $r_{max} = \opt / \norm{\one_{I_1}}$. Then, we have $b_i \leq r_{max}$ for all $i$. For those buckets $i$ with $b_i \leq r_{max}/k$, we can guess those $b_i = 0$ since the total norm of these buckets is upper bounded by $\opt$. Since $b_i$ is between $r_{max}/k$ and $r_{max}$, we guess $b_i$ up to a factor of 2 with $\log_2 k$ values $r_{max}/2^{j}$ for $j \leq \log_2 k$. Since values $\{b_i\}$ are monotone non-increasing, we guess $\{b_i\}$ with $2^{2\log_2 k} = O(k^2)$ different sequences. 
\end{proof}

\section{Hardness of Norm Multiway Cut without Oracle}\label{sec:hardness}
In this section, we prove a hardness result for the Norm Multiway Cut problem in the setting where the monotonic norm $\|\cdot\|$ is given by an explicit formula, but we have oracle access neither to a minimization nor ordering oracle.
We show that assuming the Hypergraph Dense-vs-Random Conjecture there is no $n^{1/4 - \varepsilon}$ approximation algorithm for the problem for every $\varepsilon > 0$, where $n$ is the total number of vertices.

Our proof is based on a reduction from the Small Set Bipartite Vertex Expansion (SSBVE) problem (this problem is also known as Min $k$-Union). Recall the definition of SSBVE. We are given a bipartite graph $G=(L, R, E)$ and a parameter $t$. For every $S\subseteq L$ or $S\subseteq R$, let $N(S)$ be the set of its neighbors; let $N(u) = N(\{u\})$. In SSBVE, the goal is to find a subset $S\subseteq L$ of $t$ vertices so as to minimize $|N(S)|$.

\subsection{Reduction from SSBVE to Norm Multiway Cut.} Consider an instance $G$ with parameter $t\leq k$ of SSBVE. Let $k = |L|$, $n_R = |R|$, and $n_G=k+n_R$.
To simplify the notation, we denote vertices of $L$ by $1,\dots, k$. We now define an instance $\cal I$ of Norm Multiway Cut corresponding to $G$. The instance graph is a complete bipartite graph $H = (L, B, L \times B)$ with $L = \{1,\dots,k\}$ and $B = [t] \equiv \{1,\dots, t\}$ with the set of terminals $L$ and all edges having unit weight. The norm of a vector $x\in \mathbb{R}^k$ is given by the following formula:
\[ \|x\| = \sum_{v\in R} \max_{i\in N(v)} |x_i|
\]
(where $N(v)$ is the neighborhood of $v$ in $G$). Observe that the norm is monotonic.

\begin{claim}\label{hardness-one}
Assume that $k \geq 2$. Consider an instance $\cal I$ of Norm Multiway Cut defined above. Let $(P_i)_{i\in L}$ be a feasible solution for ${\cal I}$. Denote its cost by $c$.
Define vector $y\in {\mathbb R}^{L}$ as follows. For every $i \in L$, let $y_i = |P_i| -1$ be the number of non-terminals assigned to terminal $i$. Then 
\[
(k-2) \|y\| \leq c \leq (k-2) \|y\| + t n_R.
\]
\end{claim}
\begin{proof}
Let us compute the number of edges $x_i$ leaving $P_i$:
\begin{itemize}
\item 
first, since $H$ is a complete bipartite graph, there are $(k - 1)\cdot y_i$ edges between terminals $j$ outside of $P_i$ and non-terminals in $P_i$ (namely, all edges in $(L\setminus \{i\}) \times (P_i\setminus\{i\})$
\item then there are $t-y_i$ edges between terminal $i$ and non-terminals outside of $P_i$ (namely, $\{(i, v): v\in B\setminus P_i\}$). 
\end{itemize}
We conclude that $x_i = (k-1)y_i + t-y_i = (k - 2)y_i + t$.
Let $\mathbf{1}$ be the all one vector. Note that $\|\mathbf{1}\| = \sum_{v\in R} \max_{i\in N(v)} 1 \leq \sum_{v\in R} 1 = n_R$. Write, $x = (k-2)y + t \mathbf{1}$. We have 
\begin{align}
c &= \|x\| = \|(k-2)y + t \mathbf{1}\| \geq  
(k-2) \|y\|,\\
c &= \|x\| \leq \|(k-2)y\| + t\|\mathbf{1}\| \leq  (k-2) \|y\| + t n_R ,
\end{align}
Here, we used that the norm $\|\cdot\|$ is monotonic and the triangle inequality. 
\end{proof}
\begin{claim}\label{hardness-two} Assume that there is an $\alpha = \alpha(n_H)$ approximation algorithm $\cal A$ for the Norm Multiway Cut problem (where $n_H$ is the number of vertices in the instance). Then, there is a polynomial-time algorithm that given an instance $G$ of SSBVE of cost $OPT$ finds a subset $S'\subseteq L$ of size at most $t$ such that
\[
N(S') \leq O(\alpha \log k) \frac{|S'|}{t} \Bigl(OPT + \frac{2tn_R}{k}\Bigr),
\]
where $\alpha = \alpha(n_H)$ and $n_H = k+t$.
\end{claim}
\begin{proof}
If $k \leq 3$ we find an optimal solution for SSBVE using exhaustive search, and we are done. So we assume below that $k \geq 4$.
We transform $G$ to an instance ${\cal I} = (H, \|\cdot\|)$ of Norm Multiway Cut as described above. Consider an optimal solution $S^*\subset L$ for $G$. 

First, we upper bound the cost of $\cal I$ in terms of $OPT$ and other parameters. Consider an arbitrary partition of $H$ that assigns each of $t$ non-terminals to a distinct terminal in $S^*$ (this is possible since $|S^*|=t$). Let us apply Claim 1 to compute the cost of this solution. Vector $y_i^*$ for this solution equals $1$ if $i\in S^*$ and $0$ otherwise. Further, 

\[\|y^*\| = \sum_{v\in R} \max_{i\in N(v)} |y^*_i| = 
\sum_{v\in R: N(v) \cap S^* \neq \varnothing} 1 = 
|N(S^*)| = OPT.\]

Therefore, instance $\cal I$ has cost at most $c \leq (k-2) OPT + t n_R$. We apply approximation algorithm $\cal A$ and get a solution for $\cal I$ of cost at most $\alpha c$. Consider vector $y$ for the obtained solution. Applying Claim 1 again, we get,
\[
\|y\| \leq \frac{\alpha c}{k-2} \leq \alpha \cdot \Bigl(OPT + \frac{t}{k-2} n_R\Bigr) \leq  \alpha\cdot \Bigl(OPT + \frac{2tn_R}{k}\Bigr).
\]

Note that coordinates of $y$ are integers between $0$ and $t$, and $\|y\|_1 = \sum y_i = |B| = t$.
We partition all positive coordinates of $y$ into $\ell = \lfloor\log_2 t\rfloor +1$ groups $Q_0,\dots, Q_{\ell-1}$: $Q_j = \{i: 2^j \leq y_i < 2^{j+1}\}$. Let $y^{(j)}$ be the indicator vector of $Q_j$. Then
$2^j y^{(j)} \leq y$ (coordinate-wise) and thus $\|y^{j}\| \leq \frac{1}{2^j} \|y\|$ for every $j$. Also, 
\[
\sum_{j=0}^{\ell-1} 2^j |Q_j| = \sum_{j=0}^{\ell-1} 2^j \|y^{(j)}\|_1 \geq \frac{\|y\|_1}{2} = \frac{t}{2}.
\]

Therefore, one of the summands is at least $t/(2\ell)$; that is, for some $j$, $|Q_j| \geq 2^{-(j+1)}t/\ell$. Our algorithm outputs $S' = Q_j$ for such an index $j$.
We have
\begin{align*}
&|Q_j| = \|y^{(j)}\|_1 \leq \|2^{-j} y\|_1 \leq \|y\|_1 = t,\\
&|N(Q_j)| = \|y^{(j)}\| \leq 2^{-j} \|y\| \leq 2\ell \frac{|Q_j|}{t} \|y\| \leq 2\alpha\ell \frac{|Q_j|}{t}  (OPT + 2tn_R/k).
\end{align*}
It follows that $S' = Q_j$ satisfies the requirements of the claim.
\end{proof}

\begin{claim}\label{hardness-three} Assume that there is an $\alpha = \alpha(n_H)$ approximation algorithm $\cal A$ for the Norm Multiway Cut problem (we assume that $\alpha(n_H)$ is a non-decreasing function of $n_H$). Then there is an algorithm that finds a solution for SSBVE of cost at most
\[
O(\alpha \log^2 k)  \left(OPT + \frac{n_Rt}{k-t}\right),
\]
where $OPT$ is the cost of the optimal solution.
\end{claim}
\begin{proof} Let $G = (L, R, E)$ with parameter $t$ be the input SSBVE instance.  Denote an optimal solution for $G$ by $S^*$ and its cost by $OPT$. (We use $S^*$ only in the analysis.)

We find an approximate solution by iteratively applying the algorithm from Claim \ref{hardness-two}. First, we run the algorithm on the entire instance $G_1 = G$ with $t_1 = t$ and obtain a subset $S_1 = S'$. Then we run it on subgraph $G_2 = G[(V\setminus S_1) \cup R]$ with $t_2 = t - |S_1|$, then on $G_3 = G[V\setminus(S_1\cup S_2) \cup R]$ with $t_3 = t - |S_1| - |S_2|$, and so on. Specifically, at iteration $j$, we run the algorithm on 
$G_j = G[L\setminus (S_1 \cup \dots\cup S_{j-1}) \cup R]$ with $t_j = t - |S_1| - \dots - |S_{j-1}|$.
We stop once $|S_1| + \dots + |S_{j}|\geq t$. 
Note that $|S^*\setminus (S_1\cup \dots \cup S_{i-1})| \geq |S^*| - (t - t_i) = t_i$. Thus, every subset of $S^*\setminus (S_1\cup \dots \cup S_{i-1})$ of size $t_i$ is a feasible solution for instance $G_i$ with parameter $t_i$; its cost is at most $|N(S^*)| = OPT$. By Claim~\ref{hardness-two}, 
$N(S_i) \leq O(\alpha \log k)  (OPT + 2t_i n_R/k_i) \cdot \frac{|S_i|}{t_i}$. Note that $t_i/k_i \leq t / (k -t)$.
Therefore,
\[
N\Bigl(\bigcup_{i=1}^j S_i\Bigr) \leq \sum_{i=1}^j N(S_i) \leq 
O(\alpha\log k)  \left(OPT + \frac{n_Rt}{k-t}\right) \cdot \sum_{i=1}^j\frac{|S|_i}{t_i}.\]
Now, $\frac{|S_i|}{t_i} = \frac{t_i - t_{i+1}}{t_i} \leq \log_e \frac{t_i}{t_{i+1}}$ (where we let $t_{j+1} = \sum_{i=1}^j |S_i|$); here we use that $1 - z \leq \log_e 1/z$ for $z\in (0,1]$. Therefore,
\[
N\Bigl(\bigcup_{i=1}^j S_i\Bigr) \leq 
O(\alpha\log k)  \left(OPT + \frac{n_R t}{k-t}\right) \cdot \sum_{i=1}^j\log_e \frac{t_i}{t_{i+1}} = O(\log^2 k)  \left(OPT + \frac{n_R t}{k-t}\right).
\]

Now we output an arbitrary subset $S'$ of $S_1 \cup \dots \cup S_j$ of size $t$. We have, $|N(S')| \leq |N(S_1 \cup \dots \cup S_j)| \leq O(\log^2 k)  \left(OPT + \frac{n_R t}{k-t}\right)$.
\end{proof}

We use the following corollary from the Hypergraph Dense-vs-Random Conjecture~\cite[Intro and Appendix A]{CDM17}:

\begin{corollary}[\cite{CDM17}]
For every integer $r \geq 3$ and sufficiently small $\varepsilon > 0$, there is no polynomial-time algorithm for the following task. Given an instance of SSBVE on a bipartite graph $G = (L, R, E)$ with $|L| = k$, $|R| = n_R = k^{1/2+O(\varepsilon)}$, and $t = \Theta(\sqrt{k})$, the algorithm distinguishes between the following two cases:

\begin{itemize}
\item The cost of the optimal solution is at most $k^{1/4 + O(\varepsilon)}$. 
\item The cost of the optimal solution is at least $k^{1/2 - O(1/r)}$.
\end{itemize}
\end{corollary}

Assume that there is an $\alpha = \alpha(n)$ approximation algorithm for Norm Multiway Cut. From Claim~\ref{hardness-three}, it follows that
given an instance with $OPT \leq k^{1/4 + O(\varepsilon)}$ as in the conjecture, we can find a solution of cost at most $\alpha \cdot \log^2 k \cdot (k^{1/4 + O(\varepsilon)} + n_Rt/(k-t)) \leq \alpha k^{1/4 + O(\varepsilon)} \log^2 k$.
If the conjecture holds, this number must be at least $k^{1/2 - O(1/r)}$. Thus, 
\[
\alpha(n) \geq k^{1/4 - O(\varepsilon) - O(1/r)}/\log^2 k = n_G^{1/4 - O(\varepsilon) - O(1/r)}.\]
for every $\varepsilon > 0$ and $r\geq 3$ (recall that $n_G = k + n_R = O(k)$). We have proved the following theorem.

\begin{theorem}
Assuming the Hypergraph Dense-vs-Random Conjecture, for every $\varepsilon >0$, there is no polynomial-time algorithm for Norm Multiway Cut with approximation factor $\alpha(n_G) \leq n_G^{1/4-\varepsilon}$.
\end{theorem}
Note that $k$ is polynomial in $n$ in hard instances we construct. The problem is much easier when $k$ is a constant. Indeed, then the oracles used in our algorithms can be implemented in polynomial time, and thus we can achieve polylogarithmic approximation in polynomial time.



\bibliography{references}


\end{document}